\documentclass{aa}
\usepackage{newtxtext, newtxmath}
\usepackage{bm, enumitem}
\usepackage{orcidlink}
\usepackage{xcolor}
\definecolor{darkblue}{cmyk}{55,17,0,0}
\hypersetup{
    colorlinks=true,
    linkcolor=darkblue,
    linkbordercolor=darkblue,
    citecolor=darkblue,
    urlcolor=black
}

\begin{document}
    \title{The mass-dependent interplay of active galacitc nuclei and supernova feedback
    in shaping the $L_{\rm X}$--$T$ relation of early-type galaxies}

    \author{Haojie {Xia}\inst{1,2} \orcidlink{0009-0004-3881-674X} \and Feng {Yuan}\inst{3}
    \corrauth{fyuan@fudan.edu.cn} \orcidlink{0000-0003-3564-6437} \and Bocheng {Zhu}\inst{4}
    \orcidlink{0000-0003-0900-4481} \and Haoen {Zhang}\inst{1,2} \orcidlink{0009-0008-0460-8685}
    \and Tingfang {Su}\inst{1,2} \orcidlink{0009-0003-1442-105X} \and Aoyun {He}\inst{1,2}
    \orcidlink{0000-0002-4882-8953} \and Suoqing {Ji}\inst{3} \orcidlink{0000-0001-9658-0588}
    }

    \institute{Astrophysics Division, Shanghai Astronomical Observatory, Chinese Academy of Sciences, 80 Nandan Road, Shanghai 200030, P.R.China \and University of Chinese Academy of Sciences, No. 19A Yuquan Road, Beijing 100049, P.R.China \and Center for Astronomy and Astrophysics and Department of Physics, Fudan University, Shanghai 200438, P.R.China \and National Astronomical Observatories, Chinese Academy of Sciences, 20A Datun Road, Beijing 100101, P.R.China}

    \authorrunning{H. Xia et al.}\titlerunning{Feedback in the $L_{\rm X}$--$T$ relation of ETGs}

    \date{Received: 14 April 2026 / Accepted: 18 June 2026}
    
    \abstract{The observed X-ray luminosity--temperature ($L_{\rm X}$--$T$) relation of hot gas in early-type galaxies deviates significantly from the prediction of purely gravitational heating, providing a key constraint on non-gravitational processes such as supernova (SN) and active galactic nucleus (AGN) feedback. We investigate the physical origin of this relation using high-resolution 3D hydrodynamical simulations with the multiscale AGN-regulated cosmic ecosystem resolver in 3D (MACER3D) framework, which we applied to a dwarf elliptical, a massive elliptical, and a cluster-central galaxy. For comparison, we performed controlled simulations that included AGN winds and SN feedback in isolation, excluding cosmological inflow and environmental effects. The dominant regulation mechanism depends strongly on the halo mass. In the cluster-central case, neither AGN winds nor SN feedback alone can sufficiently suppress the gas density and $L_{\rm X}$. When both are included, their nonlinear coupling suppresses the X-ray emission, producing ($L_{\rm X}$, $T$) values below the observed relation; this discrepancy can be resolved by incorporating AGN jet feedback. In massive elliptical galaxies, the inclusion of AGN feedback brings the model predictions into broad agreement with the observed $L_{\rm X}$--$T$ relation, indicating that AGN feedback dominates SN feedback.
At the low-mass end, dwarf galaxy models also follow the observed trend. In this regime, models with either SN or AGN feedback alone predict low $L_{\rm X}$. When both are included, AGN wind-driven transport of SN-enriched gas to intermediate radii enhances the metallicity and radiative cooling, thereby increasing $L_{\rm X}$. This coupled process establishes a fountain-like circulation, in which gas is repeatedly lifted and recycled within the galaxy.}

    \maketitle
    \nolinenumbers 
    \section{Introduction\label{sec:intro}}
    In the standard $\Lambda$ cold dark matter paradigm, cosmic
    structures form hierarchically through the gravitational collapse of dark
    matter halos. As primordial density fluctuations grow, baryonic gas is
    accreted into the deepening gravitational potential wells of these dark
    matter halos \citep{white78, blumenthal84}. During accretion, the gas is shock-heated
    to the virial temperature of the host halo, converting gravitational
    potential energy into thermal energy \citep{rees77, silk77}. In massive
    systems, ranging from giant elliptical galaxies to galaxy clusters, this
    virialized gas constitutes the hot interstellar medium (ISM) or intracluster
    medium (ICM), with temperatures in the range $10^{6}$--$10^{8}\,\mathrm{K}$
    \citep[e.g.,][]{mathews03, werner19}. At these temperatures, cooling
    proceeds primarily via thermal bremsstrahlung and metal line emission in the
    X-ray band \citep{sarazin86, sutherland93}. In this process, X-ray luminosity
    ($L_{\rm X}$) and spectral temperature ($T$) serve as fundamental proxies for the
    halo mass and the thermodynamic state of the hot gas, and it is therefore critical to understand the
    $L_{\rm X}$--$T$ relation for constraining models of galaxy
    formation and the thermal history of hot gaseous atmospheres \citep{borgani04, kravtsov12, allen11, giodini13}.

    To understand the physics underlying the complex observational data of these
    systems, it is essential to first establish a theoretical baseline governed solely
    by gravity. \citet{kaiser86} proposed the classic self-similar model,
    assuming that gravity is the dominant driver of structure formation and that
    the baryon fraction remains constant across the halo mass hierarchy \citep{evrard91}.
    Under the assumptions of hydrostatic equilibrium and the virial theorem, 
    a scaling relation $L_{\rm X, bol}\propto T^{2}$ is predicted \citep{bryan98}. Although
    observations in limited energy bands (e.g., soft X-rays) can slightly alter
    this slope \citep{voit05}, the fundamental theoretical expectation is a
    single continuous power law spanning from dwarf galaxies to massive clusters.

    However, X-ray observations demonstrate that real systems deviate
    significantly from this prediction, and that the deviation strongly depends on the mass.
    For galaxy clusters and groups, early studies using the Advanced Satellite for Cosmology and Astrophysics and the
    ROentgen SATellite established that the relation is steeper than predicted \citep{markevitch98, arnaud99}.
    This steepening was subsequently confirmed by high-precision observations with
    X-ray Multi-Mirror Newton, and more recently, by the extended ROentgen Survey with an Imaging Telescope Array, which enabled the
    accurate excision of cool cores. These studies established a continuous scaling
    of $L_{\rm X}\sim T^{2.5-3.5}$ across these mass regimes \citep{pratt09,lovisari15,bahar22,khalil24}.
    The steep slope implies that the ICM and the intragroup medium have been subjected
    to entropy injection or density suppression \citep{maughan12}. For example, 
    mechanical energy from active galactic nucleus (AGN) jets is now widely
    recognized as the primary heating mechanism for offsetting
    radiative cooling and preventing catastrophic gas condensation in cluster
    and group cores \citep[e.g.,][]{mcnamara07,fabian12}.

    The discrepancy becomes more pronounced in the low-mass regime of individual
    early-type galaxies (ETGs).
    This trend has been established by the detailed studies of \citet{kim15} and
    \citet{goulding16}. By combining high-mass ellipticals from the MASSIVE
    survey with lower-mass systems from the ATLAS$^{3\rm D}$ project, they
    derived a universal scaling relation characterized by a steep slope of
    $L_{\rm X}\sim T^{4.7}$ \citep[see also][]{boroson11, babyk18}. Crucially,
    this tight correlation is observed when X-ray properties are measured
    strictly within one effective radius ($r_{\rm eff}$) of the galaxy. A physically
    motivated aperture is essential for isolating the gas that directly responds to
    the galaxy potential well and for excising contamination from the extended
    ICM or large-scale group halos, which would otherwise bias measurements in
    dense environments \citep{sun07, su15, goulding16}.

    It has long been a challenge for hydrodynamic simulations to interpret the observed $L_{\rm X}$--$T$ scaling relation. Early adiabatic models failed to explain
    observations \citep{evrard90, navarro95}, demonstrating that efficient AGN
    feedback is essential to counteract radiative cooling \citep{fabian94, white91, balogh01}.
    Cosmological simulations incorporate AGN feedback using diverse sub-grid
    models.
    The limited spatial resolution necessitates phenomenological approximations. One example is the mass-accretion
    rate of the AGN. Since the Bondi radius is typically unresolved, the widely used Bondi-based prescriptions can significantly over- or underestimate the true black hole accretion rate by even several orders of magnitude \citep[e.g.,][]{hopkins16, negri17}.
    The modeling of the interaction between AGN outputs and ISM is also implemented
    as parameterized simplifications via isotropic thermal energy injection or sub-grid
    kinetic kicks, and these parameters are often treated as free.

    In this context, high-resolution galaxy-scale simulations offer clear advantages. They have been widely used to investigate the $L_{\rm X}$--$T$ relation in individual galaxies \citep[e.g.,][]{ciotti17,pellegrini25}.
We revisit this problem using the MACER framework, which was originally developed as the code called massive AGN controlled ellipticals resolved to follow the evolution of individual galaxies \citep{yuan18,yuan20}. This framework builds upon previous studies \citep[e.g.,][]{ciotti09,ostriker10,gan14}, but incorporates updated AGN feedback physics. Compared to other idealized galaxy-scale and cosmological simulations, the MACER code family has several distinctive features that are described in detail in \citet{he26}. We briefly summarize them next. First, the inner boundary of the simulation in MACER is set within the Bondi radius. By combining the mass flux measured at this boundary with black hole accretion theory, the accretion rate at the black hole horizon can be reliably estimated. Second, state-of-the-art AGN physics is implemented: the AGN outputs as functions of accretion rate (including radiation, jets, and winds) are directly informed by general relativistic magnetohydrodynamic (GRMHD) simulations and observations. Third, the interaction between these AGN outputs and the ISM is calculated self-consistently and is not parameterized through phenomenological prescriptions, as in most cosmological simulations. MACER has been successfully applied to a wide range of systems, including elliptical galaxies \citep{yuan18,yoon18,zhu23a,zhu23b}, compact galaxies \citep{di23}, dwarf galaxies \citep{su26}, disk galaxies \citep{zou26a,zou26b}, and galaxy clusters \citep{he26}. The AGN feedback physics implemented in MACER has also recently been incorporated into the cosmological simulations of IllustrisTNG \citep{Zhu26}.

The simulations in this work were based on MACER3D \citep[Multiscale AGN-regulated Cosmic Ecosystem Resolver in 3D;][]{zhang25}\footnote{\url{https://macer-project.github.io}.}. As described in \citet{zhang25}, MACER3D is the newly developed 3D version of the original MACER framework \citep{yuan18}. Beyond upgrading to the Athena++ infrastructure \citep{athena} to capture full 3D hydrodynamics, this new version includes major numerical optimizations and improved physical modules. The expansion of the acronym was updated to reflect its enhanced capability in modeling a broader range of cosmic ecosystems beyond elliptical galaxies.
    We performed simulations for three representative cases with different
    gravitational potentials: a dwarf elliptical, a massive elliptical,
    and a cluster-central galaxy hosted by a cluster-scale halo.
    We treated these systems as physical laboratories to investigate the effect
    of different feedback mechanisms. For this purpose, we performed controlled
    numerical experiments by selectively activating AGN and supernova (SN) feedback
    modules in idealized isolated environments without invoking cosmological
    inflow or environmental stripping. This choice was motivated by the observational
    evidence that the $L_{\rm X}$--$T$ relation only weakly depends
    on the environment \citep{goulding16}. Through these controlled comparisons, we wished to clarify how the interplay between AGN and stellar feedback evolves across the mass of the system, and how this mass-dependent coupling shapes the
    observed $L_{\rm X}$--$T$ relation.

    This paper is organized as follows. In Section~\ref{sec:method} we describe
    the MACER simulation setup, including the physical models for AGN and
    stellar feedback, as well as the methods used to calculate X-ray properties from
    the simulation outputs. Section~\ref{sec:results} presents the
    hydrodynamical evolution of these galaxies and analyzes their X-ray
    properties by directly comparing the results with observational scaling relations.
    In Section~\ref{sec:discussion} we discuss the physical mechanisms
    for the steepening of the $L_{\rm X}$--$T$ relation. We summarize our results
    in Section~\ref{sec:conclusion}.

    \section{Methods\label{sec:method}}
    \subsection{MACER model}
    MACER solves the time-dependent Euler equations of hydrodynamics in spherical
    polar coordinates $(r, \theta, \phi)$. The system includes source terms for
    the mass, momentum, and energy injection from the evolving stellar
    population, as well as the feedback driven by the central supermassive black
    hole (SMBH). The computational domain is discretized using a logarithmically
    spaced radial grid, which ensures a high resolution in the nuclear region. The
    grid spacing satisfies $\Delta r/r \lesssim 0.1$. The angular coverage spanned
    the full azimuthal range $\phi \in [0, 2\pi]$, while the polar angle was
    restricted to $\theta \in [5^{\circ}, 175^{\circ}]$ to mitigate the strict
    timestep constraints imposed by the Courant condition near the poles. The outer
    boundary was set to allow for free outflow, preventing artificial wave
    reflection. This setup represents an idealized isolated galaxy, in which no
    cosmological inflow or external confining medium is imposed. The inner boundary
    serves two purposes. It acts as a sink for inflowing gas to measure the
    accretion rate and as the injection boundary for AGN outflows and radiation.

    \subsubsection{AGN feedback}
    We distinguished between two accretion modes based on the black hole accretion
    rate in units of the Eddington rate,
    $\dot{m}= \dot{M}_{\rm BH}/\dot{M}_{\rm Edd}$. The Eddington accretion rate is
    defined as $\dot{M}_{\rm Edd}= L_{\rm Edd}/(\eta c^{2})$, where $\eta$ is
    the radiative efficiency, and $c$ is the speed of light. The Eddington luminosity
    is given by $L_{\rm Edd}= 4\pi G M_{\rm BH}m_{\rm p}c / \sigma_{\rm T}$,
    where $G$ is the gravitational constant, $m_{\rm p}$ is the proton mass, and
    $\sigma_{\rm T}$ is the Thomson-scattering cross-section. Following previous
    works \citep{mc06,yuan14}, we adopted a critical value of $\dot{m}_{\rm crit}
    = 0.02$ to separate the two modes.

    In the hot accretion mode ($\dot{m}< 0.02$), the accretion flow consists of
    a hot, geometrically thick inner flow and a truncated outer thin disk \citep{yuan14}.
    The truncation radius $r_{\rm tr}$ is given by \citep{yuan14}
    \begin{align}
        r_{\rm tr}\approx 3 r_{\rm s}\left[\frac{0.02\dot{M}_{\rm Edd}}{\dot{M}(r_{\rm in})}\right]^{2},
    \end{align}
    where $\dot{M}(r_{\rm in})$ is the mass-inflow rate at the inner simulation
    boundary, and $r_{\rm s}= 2GM_{\rm BH}/c^{2}$ is the Schwarzschild radius.
    The accretion rate onto the black hole is then related to the inflow rate by
    $\dot{M}_{\rm BH}\approx \dot{M}(r_{\rm in})(3r_{\rm s}/r_{\rm tr})^{0.5}$. In
    this mode, strong wide-angle winds are launched from the hot flow \citep{yuan12}.
    The mass-flow rate and velocity of these winds were directly taken from the
    3D GRMHD simulations of black hole accretion \citep{yuan15}, 
    \begin{align}
        \dot{M}_{\rm wind, hot} & \approx \dot{M}(r_{\rm in})-\dot{M}_{\rm BH}, \\
        v_{\rm wind, hot}       & = 0.2 v_{\mathrm{K}}(r_{\rm tr}),
    \end{align}
    where $v_{\mathrm{K}}$ is the Keplerian velocity. The polar angular distribution
    of the hot wind spanned $\theta \sim 30^{\circ}$-$70^{\circ}$ and $110^{\circ}$-$1
    50^{\circ}$. 

    One caveat of our simulations is that we did not include a jet. Our main motivation was to reduce the computational cost caused by the high speed of a jet. Physically, the importance of jet feedback depends on the jet power, which sensitively depends on the black hole spin. Unfortunately, the spin of a SMBH is uncertain because it depends on the accretion history of the black hole \citep{2019ApJ...873..101Z,2021ARA&A..59..117R}. In a galaxy cluster, jet feedback is widely recognized as a crucial heating source that offsets cooling  flows in galaxy cluster cores \citep[e.g.,][]{mcnamara07, fabian12}. Its impact
    on less massive individual galaxies is less clear, however. We acknowledge that during their initial  breakout phase or in gas-rich environments, jets can strongly couple with and perturb the multiphase ISM \citep[e.g.,][]{wagner12, mukherjee16, harrison18}.
    Our analysis primarily focused on the thermodynamic properties
    of the diffuse hot X-ray-emitting gas within the galactic effective radius
    ($r<r_{\rm eff}$). For a galaxy cluster, the potential well is deeper, so it is difficult for jet to pierce the ICM, and thus, a jet might strongly affect the thermodynamic structure. Specifically, \citet{he26} recently reported that the shear between jet and wind produces strong turbulence that efficiently dissipates the kinetic energy of the AGN wind and jet into heating of the ICM and suppresses the cooling flow. We discuss the effect of including a jet in Section~\ref{sec:disc_jets}. In isolated massive and dwarf
    ellipticals, however, the jet might break out of the galaxy more easily and thus deposit less energy in the ISM. Consequently,  a large fraction of its kinetic energy is
    thermalized at larger radii in the circumgalactic
    medium \citep[CGM; e.g.,][]{churazov01, vernaleo06, bourne17}. This speculation seems to have the following observational support. 
    Observationally, \citet{goulding16} found no significant correlation between
    the residuals of the $L_{\rm X}$--$T$ relation and 1.4~GHz radio power, which is a commonly
    used proxy for the jet mechanical power. This suggests that the jet heating does not
    primarily regulate the inner hot gas in typical ellipticals. This might also be the case in dwarf galaxies, whose potential well is even shallower. In addition, the crossing time of the jet 
    through the inner region might be shorter than the growth timescale of the shear-driven  Kelvin--Helmholtz instability, which also  limits the jet energy dissipation within $r\lesssim r_{\rm eff}$. Still, we keep this caveat in mind and will investigate the effect of including a jet in future works. 

    In the cold accretion mode ($\dot{m}> 0.02$), the accretion disk extends
    down to the innermost stable circular orbit. The mass flux and velocity of wind were directly taken from the observational statistical result of an AGN wind, which  scales with the bolometric luminosity, $L_{\rm BH}=\eta \dot{M}_{\rm BH}c^{2}$ where $\eta = 0.1$ \citep{gofford15} as
    \begin{align}
        \dot{M}_{\rm wind, cold} & =0.28\left(\frac{L_{\rm BH}}{10^{45}\,\mathrm{erg\,s^{-1}}}\right)^{0.85}\,\mathrm{{M_\odot}\, yr^{-1}},                             \\
        v_{\rm wind, cold}       & =\min\left(2.5\times10^{4}\left(\frac{L_{\rm BH}}{10^{45}\,\mathrm{erg\, s^{-1}}}\right)^{0.4}, 10^{5}\right)\,\mathrm{km\, s^{-1}}.
    \end{align}
    The wind velocity saturates at $\sim 10^{5}\,\mathrm{km\, s^{-1}}$ \citep{gofford15}.
    The polar angular distribution of the cold wind was assumed to follow a
    $\cos^{2}\theta$ dependence \citep{yuan18}.

    In certain evolutionary stages, the black hole can briefly enter a super-Eddington
    accretion phase ($\dot{m}\gtrsim 1$) and produce powerful outflows. The properties of the wind in this regime were taken from the general relativity radiative transfer magnetohydrodynamic simulations of \citet{2023MNRAS.523..208Y}. 

    Radiative feedback from the AGN was also incorporated in the simulations, where it appears
    as a heating and cooling term in the energy equation and as a radiation pressure
    gradient term in the momentum equation. The radiative efficiency of the accretion flow in the hot and cold modes was
    taken from \citet{xie12}, 0.1 (sub-Eddington cold mode) and \citet{jiang19} (super-Eddington cold mode), respectively. The heating and cooling term was calculated using the cooling tables generated by the photoionization code Cloudy  \citep{cloudy}, processed with the integration scheme of \citet{townsend} and
    accounting for dependences on gas density, temperature, redshift,
    metallicity, and AGN luminosity \citep{zhang25}.

    \subsubsection{Stellar feedback and star formation}
    Star formation was treated as a sink term that depletes cold, dense gas ($T < 4 \times
    10^{4}\,\mathrm{K}$, $n > 1\,\mathrm{cm^{-3}}$) from the ISM. The local star
    formation rate density is given by
    \begin{align}
        \dot{\rho}_{\rm SF}= \varepsilon_{\rm SF}\frac{\rho_{\rm gas}}{\tau_{\rm SF}},
    \end{align}
    where $\varepsilon_{\rm SF}=0.1$ is the star formation efficiency, and
    $\rho_{\rm gas}$ is the gas density. The characteristic timescale was defined
    as $\tau_{\rm SF}= \max(\tau_{\rm cool}, \tau_{\rm dyn})$, where
    $\tau_{\rm cool}$ and $\tau_{\rm dyn}$ are the
    cooling and dynamical timescales. 
    The dynamical timescale was defined as the minimum of the free-fall
    time, $\tau_{\rm ff}= \sqrt{3\pi/(32G\rho_{\rm gas})}$, and the rotational timescale,
    $\tau_{\rm rot}= r/v_{\rm c}$, where $r$ is the radius, and $v_{\rm c}$ is the
    circular velocity \citep{ciotti10, yuan18}.

    Stellar feedback replenishes the ISM with mass, metals, and thermal energy, with
    each SN releasing $E_{\rm SN}= 0.85 \times 10^{51}\,\mathrm{erg}$.
    For type Ia SNe, we adopted the  delay-time distribution from \citet{ciotti12}.
    Each event injected $M_{\rm Ia}= 1.4\,M_{\odot}$ of pure metal ejecta. Type II
    SNe occurred with a characteristic delay of $2 \times 10^{7}\,\mathrm{yr}$
    following star formation and injected an IMF-averaged ejecta mass of $M_{\rm
    II}= 16.6\,M_{\odot}$ \citep{sukhbold16}, with metallicity-dependent yields taken
    from \citet{hopkins18}.
    We accounted for stellar winds from intermediate- and low-mass stars
    ($M_{*}< 8\,M_{\odot}$). The stellar wind mass loss and chemical yields were computed by 
    interpolating the tabulated yields from \citet{nomoto13} based on the 
    instantaneous turn-off mass $M_{\mathrm{TO}}$ and stellar metallicity. For 
    the adopted stellar population age, $M_{\mathrm{TO}}\sim 1.4\,M_{\odot}$, 
    these channels provide a significant contribution from evolved stars.

    \subsubsection{Galaxy model and initial conditions\label{sec:galaxy}}
    To investigate the feedback mechanisms across a wide range of galaxy masses,
    we constructed three distinct galaxy models representing a dwarf elliptical (dE),
    a massive elliptical (mE), and a cluster-central galaxy (CCG). The
    structural parameters for these models were chosen to span the corresponding
    range of gravitational potentials and are summarized in Table
    \ref{tab:model_params}.
    \begin{table}[ht]
    \centering
    \caption{Structural parameters of the galaxy models.}
    \label{tab:model_params}
    \begin{tabular}{l c c c}
        \hline
        \hline
        Parameter                           & dE                  & mE                   & CCG                  \\
        \hline
        Stellar Component                   &                     &                      &                      \\
        \hspace{1em} Profile Type           & Plummer             & Jaffe                & Jaffe                \\
        \hspace{1em} $M_{\star}$ [$M_{\odot}$]           & $1.5 \times 10^{9}$ & $3.0 \times 10^{11}$ & $1.2 \times 10^{12}$ \\
        \hspace{1em} $r_{\rm p}$ / $r_{\rm J}$ [kpc]     & 0.3                 & 9.3                  & 22.0                 \\
        \hspace{1em} $r_{\rm eff}$ [kpc]                 & 0.3                 & 6.9                  & 16.8                 \\
        \hspace{1em} $\sigma_{0}$ [km s$^{-1}$]          & 70                  & 260                  & 396                  \\
        \hline
        Dark Matter Halo                   &                     &                      &                      \\
        \hspace{1em} Profile Type          & NFW                 & SIS                  & SIS                  \\
        \hspace{1em} $v_{\rm c}$ [km s$^{-1}$]           & --                  & 368                  & 900                  \\
        \hspace{1em} $\rho_{s}$ [$M_{\odot}\,\mathrm{kpc}^{-3}$] & $8.0 \times 10^{6}$ & --                   & --                   \\
        \hspace{1em} $r_{s}$ [kpc]                       & 7.55                & --                   & --                   \\
        \hline
        Black Hole                         &                     &                      &                      \\
        \hspace{1em} $M_{\rm BH}$ [$M_{\odot}$]          & $9.0 \times 10^{5}$ & $1.8 \times 10^{9}$  & $7.2 \times 10^{9}$  \\
        \hline
        Gas                                &                     &                      &                      \\
        \hspace{1em} $n_{0}$ [cm$^{-3}$]                 & 0.1                 & 0.08                 & 0.08                 \\
        \hspace{1em} $r_{\rm c}$ [kpc]                   & 2.5                 & 6.9                  & 16.8                 \\
        \hspace{1em} $Z_{0}$ [$Z_{\odot}$]               & 0.1                 & 2.0                  & 2.0                  \\
        \hline
        Simulation Domain                  &                     &                      &                      \\
        \hspace{1em} $r_{\rm in}$ [pc]                   & 7                   & 25                   & 25                   \\
        \hspace{1em} $r_{\rm out}$ [kpc]                 & 75                  & 250                  & 500                  \\
        \hline
    \end{tabular}
\end{table}

    For the mE and CCG models, we adopted structural properties typical of high-mass
    ETGs. For simplicity, these idealized models were assumed to be spherically symmetric and nonrotating. The stellar density profile, $\rho_{\star}(r)$, followed the Jaffe
    profile \citep{jaffe83}, 
    \begin{align}
        \rho_{\star}(r)=\frac{M_{\star}r_{\rm J}}{4\pi r^{2}(r_{\rm J}+r)^{2}},
    \end{align}
    where $M_{\star}$ is the total stellar mass, and $r_{\rm J}$ is the
    characteristic radius. For the CCG, we set
    $M_{\star}= 1.2 \times 10^{12}\,M_{\odot}$ and $r_{\rm J}= 22\,\mathrm{kpc}$, yielding an
    effective radius of $r_{\rm eff}= 16.8\,\mathrm{kpc}$. We note that this CCG model represents a system with an extremely high mass, analogous to the brightest cluster galaxy in a rich cluster. This was intentionally designed to test the feedback coupling in an extremely deep potential well. The mE model represents a lower-mass  analog, with $M_{\star}= 3 \times 10^{11}\,M_{\odot}$ and $r_{\rm J}= 9.3\,\mathrm{kpc}$  ($r_{\rm eff}= 6.9\,\mathrm{kpc}$). 
    The gravitational potential was dominated by the combined contribution of the stellar component and a massive dark matter  halo. For these high-mass systems, we modeled the dark matter halo as a
    singular isothermal sphere (SIS), characterized by a constant circular velocity
    $v_{\rm c}$ \citep[see, e.g.,][for detailed properties of such models]{ciotti12, gan14}. This provided a reasonable approximation to the inner regions
    of massive halos. We adopted $v_{\rm c}= 900\,\mathrm{km\,s^{-1}}$ for the CCG and
    $v_{\rm c}= 368\,\mathrm{km\,s^{-1}}$ for the mE galaxy, corresponding to virial halo
    masses of $M_{200}\approx 2.4 \times 10^{14}\,M_{\odot}$ and
    $1.6 \times 10^{13}\,M_{\odot}$, respectively. These parameters were chosen
    such that the resulting potential depth was consistent with the central
    velocity dispersions ($\sigma_{0}$) observed in such systems \citep{faber76}.

    For the low-mass dE model, we adopted the structural profiles to reflect the
    cored light distributions and dark-matter-dominated nature typical of dwarf
    systems. The stellar component followed a Plummer profile \citep{plummer11}, 
    \begin{align}
        \rho_{\star}(r) = \left(\frac{3M_{\star}}{4\pi r_{\rm p}^{3}}\right) \left(1 + \frac{r^{2}}{r_{\rm p}^{2}}\right)^{-5/2},
    \end{align}
    with a stellar mass of $M_{\star}= 1.5 \times 10^{9}\,M_{\odot}$ and a scale radius
    $r_{\rm p}= 0.3\,\mathrm{kpc}$ (equivalent to $r_{\rm eff}$). Unlike the isothermal
    halos adopted for the massive galaxies, the dark matter halo of the dE was modeled
    using a cosmologically motivated Navarro-Frenk-White (NFW) profile \citep{navarro97} as 
    \begin{align}
        \rho_{\rm DM}(r) = \frac{\rho_{s}}{(r/r_{s})(1+r/r_{s})^{2}},
    \end{align}
    where we adopted a characteristic density $\rho_{s}= 8.0 \times 10^{6}\,M_{\odot}\,\mathrm{kpc^{-3}}$ and a scale radius $r_{s}= 7.55\,\mathrm{kpc}$, corresponding to a halo mass of
    $6.4\times10^{10}\,M_{\odot}$.

    In all three models, the central SMBH was initialized
    according to the local $M_{\rm BH}-\sigma$ relation \citep{kormendy13},
    resulting in black hole masses ranging from $9 \times 10^{5}\,M_{\odot}$ (dE)
    to $7.2 \times 10^{9}\,M_{\odot}$ (CCG). The initial ISM was set as a hot
    single-phase plasma in hydrostatic equilibrium within the total
    gravitational potential. The gas number density profile followed a beta model
    with $\beta = 2/3$ as
    \begin{align}
        n(r) = n_{0}\left(1 + \frac{r^{2}}{r_{\rm c}^{2}}\right)^{-\frac{3\beta}{2}}.
    \end{align}
    The core radius $r_{\rm c}$ was set to match the stellar effective radius for
    the massive systems ($r_{\rm c}\approx r_{\rm eff}$), while for the dE, we
    adopted a larger core relative to the stars ($r_{\rm c}= 2.5\,\mathrm{kpc}$) to
    represent a more extended gas distribution. The chemical composition of the initial
    hot halo was chosen to be consistent with the typical enrichment history of galaxies
    at each mass scale. We assumed a radially dependent metallicity profile, $Z(r
    )$, which was uniform within the core ($r \le 0.125\,r_{\rm eff}$) and
    followed a power-law decline at larger radii. The central metallicity,
    $Z_{0}$, was chosen to match observational constraints: for the massive
    systems (mE and CCG), we adopted a supersolar value of $Z_{0}= 2.0\,Z_{\odot}$,
    consistent with the metal-rich cores of massive ellipticals. In contrast,
    for the dE, we adopted a subsolar value of $Z_{0}= 0.1\,Z_{\odot}$,
    reflecting the lower star formation efficiency and shallower potential wells
    of dwarf galaxies.

    The computational domain was tailored to the physical scale of each system.
    The inner boundary, $r_{\rm in}$, was placed within the Bondi radius. The outer
    boundary, $r_{\rm out}$, extended sufficiently far to encompass the bulk of the
    hot atmosphere. It also captured the gravitational potential well
    characteristic of each galaxy type to minimize numerical boundary artifacts.
    The specific dimensions of the simulation grid for each model are detailed in
    Table \ref{tab:model_params}.

    \subsection{Calculation of the X-ray properties}
    To compare our results with observations, we computed the synthetic X-ray
    luminosity ($L_{\rm X}$) and temperature ($T_{\rm spec}$) from the
    simulation outputs using a forward spectral modeling approach. We utilized the
    \textsc{AtomDB} atomic database (v3.1.3; \citealt{foster12}) via the \textsc{PyAtomDB}
    package to generate emission spectra, assuming the gas was in collisional
    ionization equilibrium (CIE). For each computational cell with volume $\mathrm{d}
    V$, the intrinsic X-ray luminosity in a given energy band was calculated as
    \begin{align}
        \mathrm{d}L_{{\rm X}}= n_{\rm e}n_{\rm H}\Lambda(T, Z) \mathrm{d}V,
    \end{align}
    where $n_{\rm e}$ and $n_{\rm H}$ are the electron and hydrogen number
    densities, respectively, and $\Lambda(T, Z)$ is the band-limited cooling
    function, including the continuum and line emission. It was computed using
    the \texttt{APEC} plasma model based on the \textsc{AtomDB} database under
    the assumption of CIE and was integrated over the specified energy band. The
    total synthetic spectrum of the galaxy was obtained by summing the emission
    from all gas cells within the computational domain (or a specified aperture,
    e.g., $r < r_{\rm eff}$).

    To derive a characteristic temperature consistent with that obtained from observational
    spectral fitting, we avoided simple mass- or volume-weighted averages, which are
    known to introduce biases in multiphase media \citep{mazzotta04}. Instead,
    we performed a single-temperature spectral fit to the synthesized spectrum. We
    constructed a grid of template spectra using the \texttt{APEC} plasma model,
    spanning the relevant range of temperatures and metallicities, and
    determined the best-fit model by minimizing the $\chi^{2}$ statistic. The resulting
    best-fit parameter was adopted as the spectroscopic temperature ($T_{\rm
    spec}$).

    We note that our spectral fitting was performed in the intrinsic spectral space
    without folding through any instrumental response, effectively corresponding
    to an idealized detector with uniform sensitivity across the energy band. While
    observational measurements depend on the instrument, this difference is
    expected to be subdominant compared to the intrinsic physical variations. This
    approximation effectively isolates the intrinsic thermal structure of the gas
    without introducing instrument-specific weighting. Finally, the total X-ray luminosity ($L_{\rm X}$) is quoted in the energy
    band of $0.3$--$5.0\,\mathrm{keV}$, matching that adopted in \citet{goulding16}.

    \section{Results\label{sec:results}}
    \subsection{Simulation suite}
    To isolate the roles of different feedback mechanisms, we ran three
    simulations for each galaxy model defined in Section~\ref{sec:galaxy}. The configurations
    are summarized in Table \ref{tab:sim_suite}. 
    \begin{table}[h]
        \centering
        \caption{Summary of the simulation suite.}
        \label{tab:sim_suite}
        \begin{tabular}{l c c}
            \hline
            \hline
            Model             & AGN Feedback & SN Feedback \\
            \hline
            \texttt{Fiducial} & On           & On                  \\
            \texttt{noAGN}    & Off          & On                  \\
            \texttt{noSN}     & On           & Off                 \\
            \hline
        \end{tabular}
    \end{table}
    
    The \texttt{Fiducial} model served as the baseline and included AGN and
    stellar feedback. The \texttt{noAGN} model turned AGN feedback off and only left
    stellar feedback to regulate the ISM. Conversely, the \texttt{noSN} model
    turned SN feedback of (both Type Ia and Type II), allowing 
    us to assess the role of AGN feedback in the absence of SN feedback. 
    Stellar winds from evolved stars remained active as a separate mass- 
    and metal-return channel.

    The simulations for the massive systems (mE and CCG) were evolved for a total
    duration of 1\,Gyr. This time span covers several cooling times of the hot
    halo and allowed the simulation to develop multiple self-regulated AGN heating-cooling cycles. In contrast, the dE models were simulated for $200\,\mathrm{Myr}$. Given the compact
    scale of the dwarf galaxy, this duration is nearly two orders of magnitude
    longer than the local dynamical timescale ($\tau_{\rm dyn}\sim 3$\,Myr). Consequently,
    this duration was sufficient for the system to evolve over many dynamical
    times, allowing us to determine its thermodynamic outcome.

    The initial redshift $z\sim2$ was adopted solely to set the stellar 
    population age for computing stellar mass loss and Type Ia SN rates. 
    Our setup was idealized. The dark matter halo was static, and cosmological 
    accretion and mergers were not included, so the redshift carried no 
    cosmological meaning. The initial gas density was deliberately set slightly low, with 
    the intention that stellar mass loss gradually built up the gas reservoir 
    and metallicity to a self-consistent quasi-steady state. This ensured that 
    our results are not biased by the initial conditions. A moderate stellar 
    feedback level was chosen to achieve this state within a feasible simulation 
    time. An old ($\sim13\,\mathrm{Gyr}$) population would make the gas supply too slow 
    and is not suitable for this purpose. The younger age overestimates the stellar 
    wind and Type Ia SN rates relative to $z\sim0$ galaxies 
    (Section~\ref{sec:disc_caveats}).
    
    \subsection{Morphological properties}
    \begin{figure*}
        \centering
        \includegraphics[width=\textwidth]{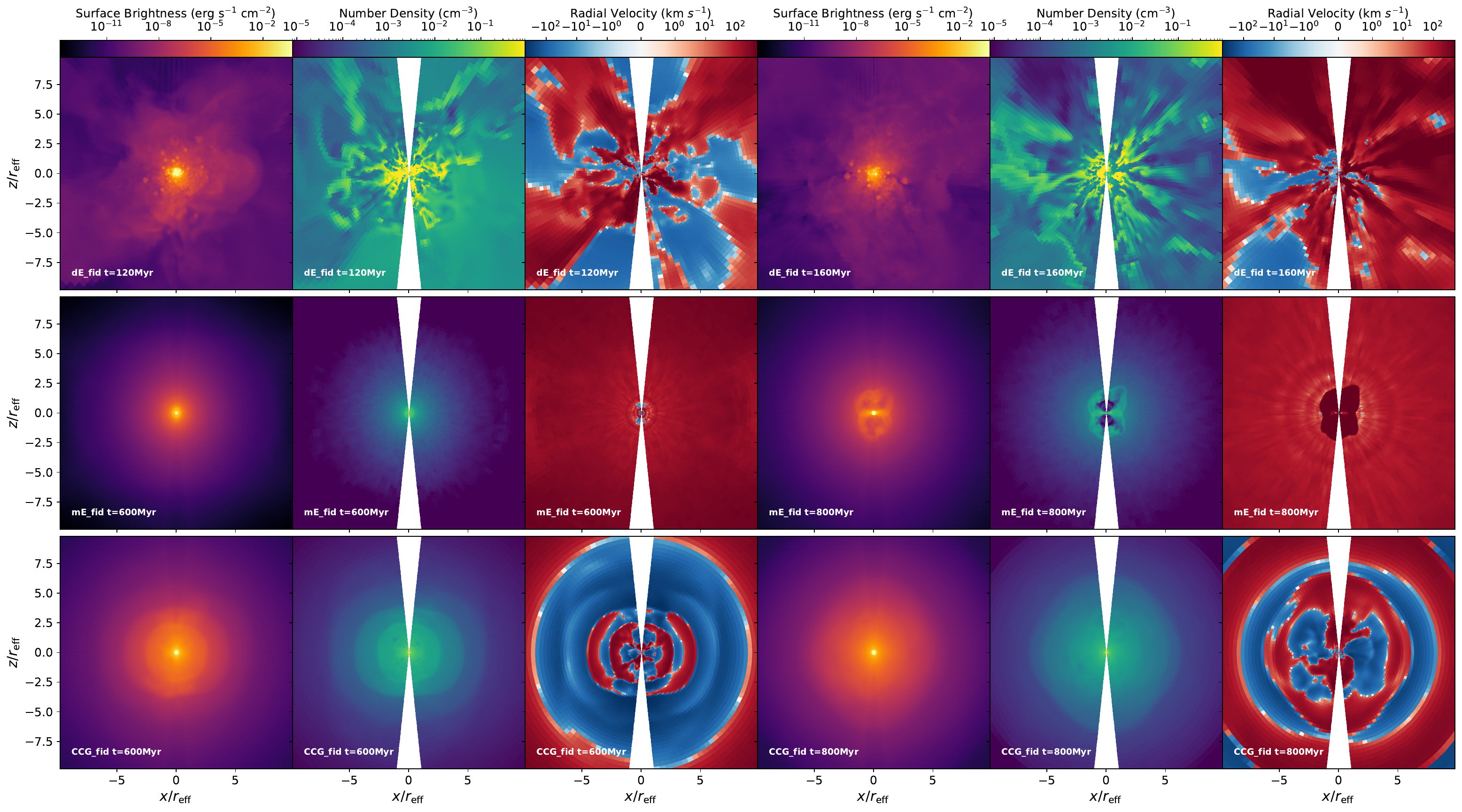}
        \caption{Morphological evolution of the gas in the fiducial runs. The rows show
        the dE (top; $t=120$ and $160\,\mathrm{Myr}$), mE (middle; $t=600$ and $800\,\mathrm{Myr}$),
        and CCG (bottom; $t=600$ and $800\,\mathrm{Myr}$) models. The columns from left to right show the
        synthetic X-ray surface-brightness projection, the midplane gas number density,
        and the radial velocity $v_{r}$ (red shows outflows, and blue shows inflows). The spatial
        coordinates are normalized by $r_{\rm eff}$. The white wedges along the polar
        axis mark the computational exclusion zone.}
        \label{fig:morphology}
    \end{figure*}
    Fig.~\ref{fig:morphology} illustrates the morphological and kinematic
    structure of the hot halo in the three fiducial runs. By combining synthetic
    X-ray surface-brightness projections with midplane slices of the gas density and
    radial velocity, we found qualitatively distinct feedback coupling patterns
    across the mass spectrum. To isolate the morphological and kinematic effects of individual feedback channels, we also present the corresponding snapshots for the \texttt{noAGN} and \texttt{noSN} control runs in Appendix~\ref{app:control_morphology}.

    The top row of Fig.~\ref{fig:morphology} shows the X-ray surface-brightness, the midplane gas number density, and the radial velocity of the dE at $t=120\,\mathrm{Myr}$ and $160\,\mathrm{Myr}$. While the density slices exhibit a disrupted central
    structure, the radial-velocity field is dominated by disordered patchy
    outflows (red) and inflows (blue) with little coherent large-scale symmetry.
    This behavior is consistent with turbulence driven by stochastic SN heating, which hinders the gas from settling into a stratified atmosphere and
    obscures coherent AGN-driven signatures.

    In contrast, the mE model (middle row) exhibits a cyclic cooling-heating feedback
    loop. At $t=600\,\mathrm{Myr}$, the system is in a cooling-dominated phase with a bright
    X-ray core and predominantly inward radial motions near the center. The
    enhanced central cooling fuels SMBH accretion and leads to an active phase
    by $t=800\,\mathrm{Myr}$. The density slice then reveals a pair of low-density cavities
    that coincide with high-velocity bipolar outflows in the radial velocity map,
    indicating that mechanical feedback from the AGN wind inflates cavities and
    displaces the surrounding gas. The resulting structures are reminiscent of the
    observed cavities and shocks in systems such as NGC~5813 \citep{randall11, randall15}.

    The bottom row shows the CCG case. The flow remains globally confined here,
    and the most prominent features in the radial velocity maps are concentric nearly
    spherical shells that propagate outward, consistent with compressive waves
    or weak shocks. In the deep potential well of the CCG, the dense and high-pressure
    ambient hot atmosphere can inhibit large-scale bulk expansion of the heated
    core. As a result, a substantial fraction of the injected energy is
    redistributed through the core via outward-propagating wave-like
    perturbations rather than being converted into sustained bulk outflows. Compared
    to the bipolar-cavity morphology in the mE and the more disordered dE flow,
    this wave-mediated coupling provides a more volume-filling quasi-isotropic channel
    for the energy transport, which might help offset radiative losses and maintain a
    quasi-hydrostatic core without requiring large-scale mass ejection.
    
    \subsection{X-ray scaling relations}
    \begin{figure}
        \centering
        \includegraphics[width=0.9\columnwidth]{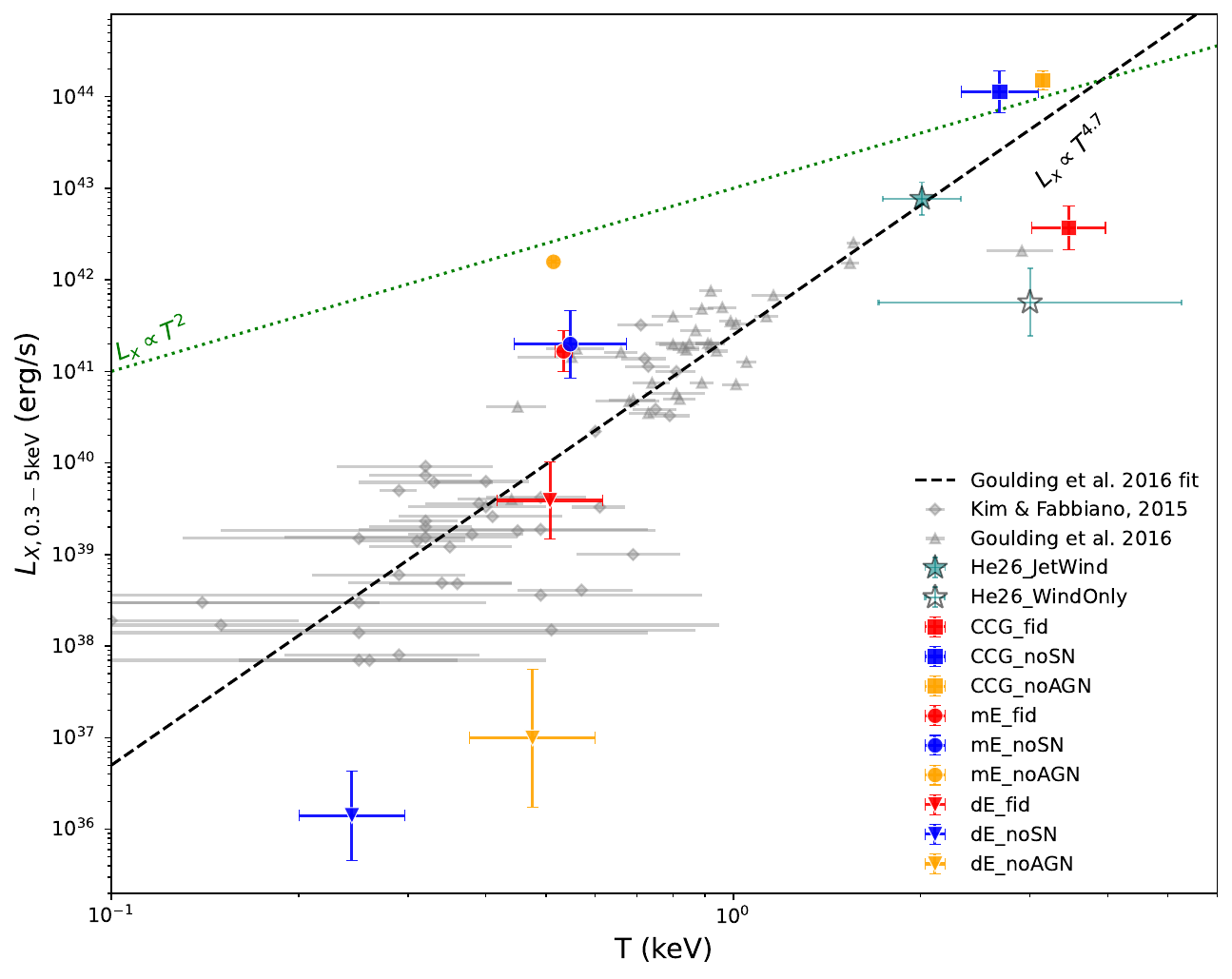}
        \caption{Mean X-ray luminosity ($L_{\rm X}$) vs. spectroscopic
        temperature ($T_{\rm spec}$) relation for the simulated galaxies within
        one effective radius ($r_{\rm eff}$). The data points represent the time-averaged
        values calculated over the latter half of the simulation to ensure the systems
        reached a quasi-steady state. The error bars denote the $1\,\sigma$
        temporal scatter, illustrating the variability of the feedback cycle rather
        than statistical fitting errors. The gray symbols and dashed line
        represent the observed samples \citep{kim15,goulding16} and the best-fit scaling relation ($L_{\rm
        X}\propto T_{\rm spec}^{4.7}$) from \citet{goulding16}. The colors
        distinguish the feedback models: \texttt{Fiducial} (red), \texttt{noAGN}
        (orange), and \texttt{noSN} (blue). The different geometric symbols denote the
        galaxy mass models: dE (triangles), mE (circles), and CCG (squares). The
        stars show the cluster simulations from \citet{he26}, with the open
        star denoting their \texttt{WindOnly} run, and the solid star denoting
        their \texttt{JetWind} run.}
        \label{fig:lx_t_relation}
    \end{figure}

Fig.~\ref{fig:lx_t_relation} shows the time-averaged X-ray luminosity and spectroscopic temperature measured within one effective radius ($r_{\rm eff}$) for our simulated galaxies, together with a comparison to observations. Overall, the \texttt{Fiducial} models broadly follow the observed $L_{\rm X}$--$T$ trend reported by \citet{goulding16}, with a notable deviation in the CCG case. The CCG model lies beyond the mass range of the observational ETG samples plotted here. We included it to theoretically investigate the feedback coupling and the resulting scaling relations in the deep potential wells of cluster-central environments. Specifically, the \texttt{CCG\_Fid} run lies below the best-fit scaling relation. In contrast, the two control runs (\texttt{noAGN} and \texttt{noSN}) produce $L_{\rm X}$ values that are higher by nearly two orders of magnitude within $r_{\rm eff}$. This comparison indicates that the full feedback model significantly suppresses X-ray emission in the central region. On the other hand, the control runs lie closer to the self-similar scaling relation, $L_{\rm X} \propto T^{2}$.

To further investigate the origin of the CCG offset, we overplot the cluster simulations from \citet{he26}, shown as stars. Their \texttt{WindOnly} run (open star) also falls below the observed relation, consistent with our \texttt{CCG\_fid} model, whereas their \texttt{WindJet} run (filled star) lies along the observational relation. The descriptions of the simulations of \citet{he26} and the physical interpretation of these results are discussed in Section~\ref{sec:disc_jets}.

In the intermediate-mass mE case, the time-averaged temperatures of all
 three runs are broadly similar, suggesting that $T$ is primarily set by the depth of the host potential. In contrast, the X-ray luminosities differ substantially.  While the \texttt{noAGN} run lies well above the observed relation, the
    \texttt{Fiducial} and \texttt{noSN} runs suppress the emission and are
    consistent with the observed scaling, implying that AGN feedback instead of SN feedback primarily drives
    the luminosity reduction. 

The observational data shown in Fig.~\ref{fig:lx_t_relation} are based on samples of massive and intermediate-mass ETGs, and do not include dwarf ellipticals \citep[e.g.,][]{kim15,goulding16}. The results for the dE case presented in the figure therefore represent a testable prediction for how feedback regulates the X-ray-emitting gas in the low-mass regime. At the low-mass end, the \texttt{Fiducial} model prediction lies close to the extrapolated observational relation, whereas the control runs exhibit significantly reduced X-ray emission. 
Specifically, the \texttt{dE\_noAGN} run shows a suppressed luminosity, while in the \texttt{dE\_noSN} run, the luminosity and temperature are substantially reduced. The prediction of the \texttt{Fiducial} model can thus be tested by future observations.

    \subsection{Physical interpretation\label{sec:physical_drivers}}

    \begin{figure*}
        \centering
        \includegraphics[width=0.9\textwidth]{
            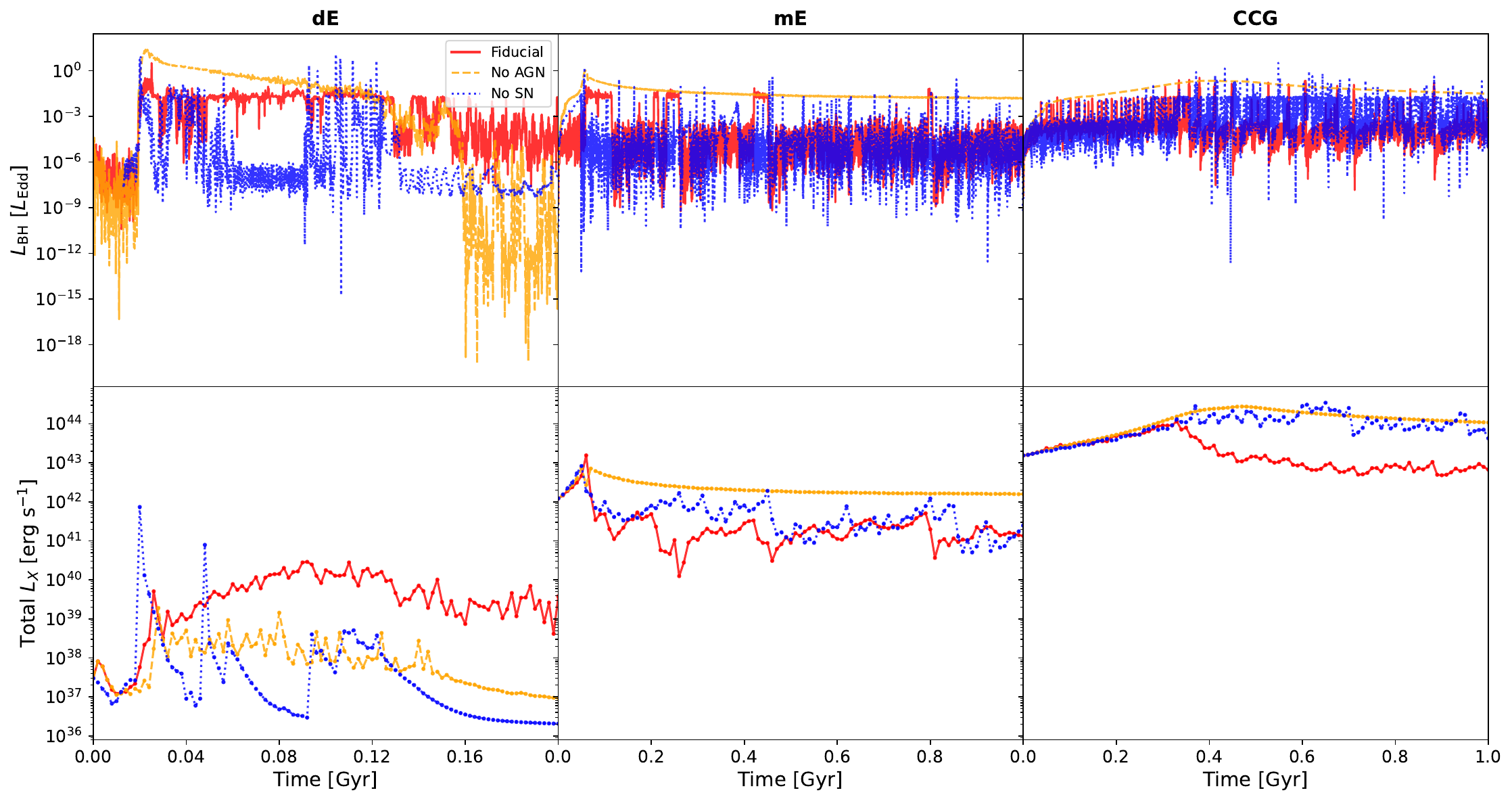
        }
        \caption{Global evolution of AGN activity and X-ray output in the dE (left),
        mE (middle), and CCG (right) models. The rows from top to bottom show the luminosity
        Eddington ratio ($L_{\rm BH}/L_{\rm Edd}$) and the total X-ray luminosity
        in the $0.3$--$5.0\,\mathrm{keV}$ band ($L_{\rm X}$). Three feedback scenarios are
        compared: the \texttt{Fiducial} run (solid red lines), the \texttt{noAGN}
        run (dashed orange lines), and the \texttt{noSN} run (dotted blue
        lines). In the \texttt{noAGN} run, $L_{\rm BH}$ is plotted only as a
        diagnostic proxy for the SMBH accretion rate and does not imply that
        AGN radiative feedback is enabled. The
        time baselines for the dwarf system ($0.2\,\mathrm{Gyr}$) are different from that of the massive systems
        ($1.0\,\mathrm{Gyr}$).}
        \label{fig:lightcurves}
    \end{figure*}

    \begin{figure*}
        \centering
        \includegraphics[width=0.9\textwidth]{
            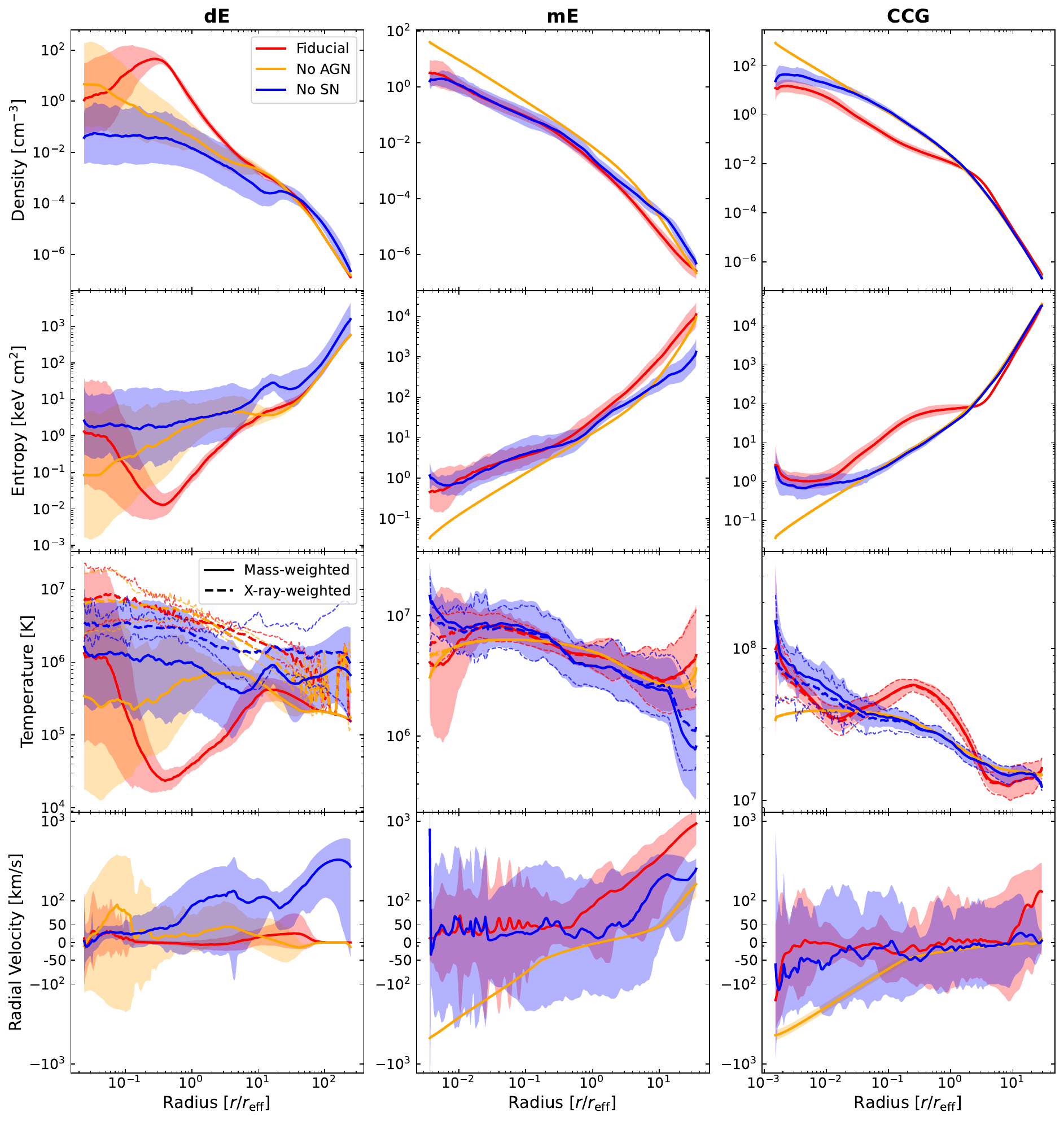
        }
        \caption{Time-averaged radial profiles of the gas properties for dE (left),
        mE (middle), and CCG (right) systems. The rows from top to bottom display the
        volume-averaged number density, the mass-weighted entropy, the gas temperature
        (both $T_{\rm mw}$ and $T_{\rm X}$; see below), and the mass-weighted radial
        velocity. The colors denote the simulation runs: \texttt{Fiducial} (red), \texttt{noAGN}
        (orange), and \texttt{noSN} (blue). All statistics were calculated over the
        latter half of the simulation period. The shaded regions indicate the
        $1\,\sigma$ temporal scatter for quantities plotted with solid curves. In
        the temperature panels, the solid curve shows the mass-weighted
        temperature ($T_{\rm mw}$), and the thick dashed curve shows the \textsc{AtomDB}-derived
        X-ray luminosity-weighted temperature ($T_{\rm X}$); the thin dashed curves
        indicate the $\pm1\,\sigma$ temporal deviations around $T_{\rm X}$.}
        \label{fig:profiles}
    \end{figure*}
    To understand these results, we examined the
    temporal evolution (Fig.~\ref{fig:lightcurves}) and the radial structure (Fig.~\ref{fig:profiles})
    of the gas jointly. This combined analysis provided insight into how the
    interplay between AGN and SN feedback depends on the gravitational
    potential depth.

    \subsubsection{The cluster-central galaxy\label{sec:CCG}}
    To interpret the CCG results, we began with the \texttt{noAGN} model. Although this run included distributed heating from SNe, radiative cooling dominated the overall energy balance. As the gas loses thermal pressure support, the initial hydrostatic equilibrium breaks down, leading to the development of a sustained radial inflow with velocities reaching about $-300\,\mathrm{km\,s^{-1}}$ throughout the inner ISM ($r \lesssim r_{\rm eff}$; Fig.~\ref{fig:profiles}, bottom right). This widespread coherent inflow is shown as the large-scale blue regions of the radial velocity map for the \texttt{noAGN} model (Fig.~\ref{fig:morphology_noAGN}, bottom row). Because the inflow timescale remains shorter than the local cooling timescale, the system enters an advection-dominated regime (Fig.~\ref{fig:profiles}, top right). This inflow maintains a steep central density cusp that substantially enhances the X-ray emissivity ($\epsilon \propto \rho^{2}$) and sustains a persistently luminous hot atmosphere.

    For the \texttt{noSN} run, AGN wind feedback alone is insufficient to offset radiative cooling on galactic scales. Radiative cooling drives a sustained inflow that fuels SMBH accretion (Fig.~\ref{fig:lightcurves}, top right), which in turn launches AGN winds. However, the impact of these winds is largely confined to the innermost region, where they excavate a central cavity ($r \lesssim 0.02\,r_{\rm eff}$). Consequently, the \texttt{noSN} model maintains a higher central gas density than the \texttt{Fiducial} run, as displayed in the corresponding density map (Fig.~\ref{fig:morphology_noSN}, bottom row). Fig.~\ref{fig:profiles} shows that the winds are not strong enough to suppress the gas density on larger scales. 
As a result, gas accumulates at intermediate radii, leading to densities that are comparable to or even exceed those in the \texttt{noAGN} model. Consequently, the total X-ray luminosity remains similar to that of the \texttt{noAGN} run. The temperature at intermediate and large radii is lower than in the \texttt{noAGN} case owing to the absence of distributed SN heating.

    In the \texttt{Fiducial} run, the combination of AGN and SN feedback fundamentally
    changes the thermodynamic structure of the ISM. By suppressing the rapid
    inflow, the AGN wind increases the inflow timescale,
    and the distributed SN heating can couple more effectively to the hot
    phase. This distributed energy input provides extended pressure support and drives
    a global expansion of the atmosphere. Consistent with this picture, the \texttt{Fiducial}
    run shows a lower gas density within $r_{\rm eff}$ and a relatively higher
    density beyond that radius (Fig.~\ref{fig:profiles}, top right). The reduced
    density within $r_{\rm eff}$ suppresses $L_{\rm X}$ by more than ten times compared to the cases that included AGN or SN feedback alone, indicating that the combined  effect of AGN and SN feedback is nonlinear and is more important in suppressing luminosity than the simple sum of AGN and SN feedback.
    
    \subsubsection{The massive elliptical\label{sec:mE}}
    In the mE case, the gravitational potential is shallower than in the CCG, which results in a substantially lower gas temperature (Fig.~\ref{fig:profiles}). At these temperatures, radiative cooling is more efficient, and the system reaches a steady state from the initial conditions on a shorter timescale. In the \texttt{noAGN} model, radiative cooling dominates within the dense inner ISM despite the distributed SN heating. The gas settles into a stable continuous cooling flow that maintains a concentrated central density cusp (Fig.~\ref{fig:profiles}, top middle). This directly corresponds to the highly luminous smooth spherical core observed in its X-ray projection (Fig.~\ref{fig:morphology_noAGN}, middle row), yielding the highest time-averaged $L_{\rm X}$ among the three models.

The inclusion of AGN feedback fundamentally alters this behavior. The \texttt{noSN} and \texttt{Fiducial} models both suppress the time-averaged X-ray luminosity by more than an order of magnitude, bringing the predictions into agreement with the observed scaling relations. Rather than completely shutting off cooling, these AGN-active models exhibit episodic cooling events that fuel SMBH accretion and trigger intermittent AGN outbursts (Fig.~\ref{fig:lightcurves}, top middle). The strong AGN winds that are launched during these outbursts displace the innermost gas, producing a flatter density profile (Fig.~\ref{fig:profiles}, top middle) and substantially reducing $L_{\rm X}$. 

    \subsubsection{The dwarf elliptical\label{sec:dE}}
    \begin{figure}
        \centering
        \includegraphics[width=0.9\columnwidth]{
            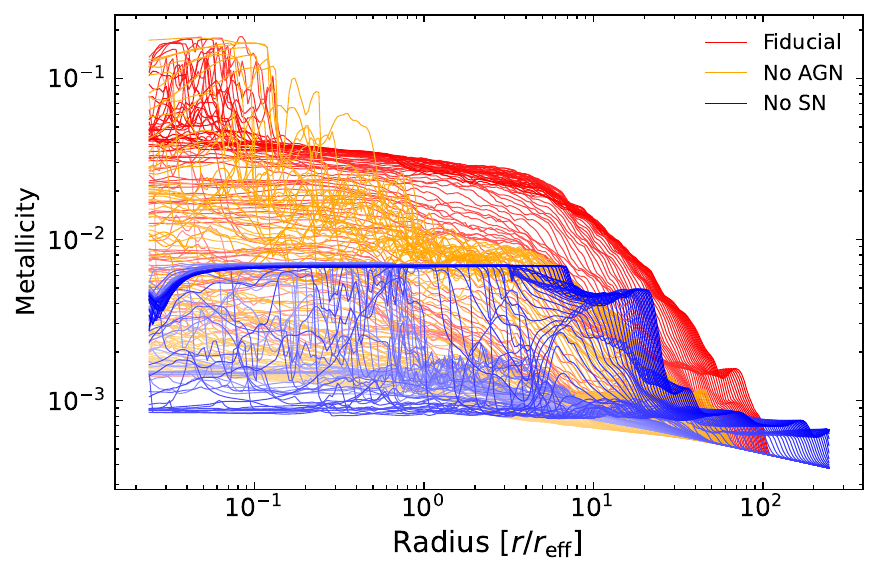
        }
        \caption{Radial evolution of the gas metallicity for the dE galaxy. The colors denote
        the simulation runs: \texttt{Fiducial} (red), \texttt{noAGN} (orange),
        and \texttt{noSN} (blue). The line opacity indicates the simulation time, ranging
        from $t=20\,\mathrm{Myr}$ (light) to $t=200\,\mathrm{Myr}$(dark).}
        \label{fig:metal_evolution}
    \end{figure}
    
    In such a shallow gravitational potential, SN heating dominates the energy budget of the hot ISM. Consequently, the \texttt{noAGN} model develops a time-averaged net outflow ($v_{r}>0$; Fig.~\ref{fig:profiles}, bottom left), showing disordered outflows driven by SN heating (Fig.~\ref{fig:morphology_noAGN}, top row) that expands the inner atmosphere and lowers the core density. This SN-driven wind produces a clear break in the density and entropy profiles at $\sim 10\,r_{\rm eff}$ (Fig.~\ref{fig:profiles}, left column). The resulting reduced core emissivity within this radius suppresses the X-ray luminosity, placing the \texttt{noAGN} model well below the self-similar expectation of $L_{\rm X}\propto T^{2}$.

The behavior of the \texttt{noSN} run is markedly different. Fig.~\ref{fig:lightcurves} shows that the AGN luminosity and $L_{\rm X}$ both vary strongly. This variability arises from a feedback cycle: when the gas cools, it efficiently fuels SMBH accretion, triggering strong AGN activity. The resulting winds expel a significant fraction of the gas to large radii, as evidenced by the progressive metal enrichment in the outskirts ($r \gtrsim 20\,r_{\rm eff}$; Fig.~\ref{fig:metal_evolution}) and by a more strongly suppressed inner density profile than in the \texttt{noAGN} run (Fig.~\ref{fig:profiles}, top left). As the gas reservoir is depleted, the accretion rate onto the black hole declines, weakening the AGN feedback. Radiative cooling then allows the gas to accumulate again, increasing the density and leading to enhanced $L_{\rm X}$ and AGN luminosity. This cyclic behavior produces large oscillations in both quantities. Overall, however, the time-averaged $L_{\rm X}$ remains significantly lower than in the \texttt{noAGN} model, as shown in Fig.~\ref{fig:lx_t_relation}.

It is noteworthy that the \texttt{Fiducial} run is the most X-ray luminous dE model although it includes both feedback channesl. This behavior arises from the coupled effects of metal enrichment, AGN wind–driven transport, and radiative cooling. Supernovae play a dual role by heating the ISM and enriching it with metals. The distributed heating increases the cooling time and stabilizes black hole accretion. On the other hand, the moderately powered AGN wind is insufficient to expel gas from the galaxy, but instead lifts metal-enriched gas from the central region to intermediate radii.

As a result, a high-metallicity plateau develops at $r \sim 0.1$--$10\,r_{\rm eff}$ (Fig.~\ref{fig:metal_evolution}). The elevated metallicity enhances the cooling function $\Lambda(T,Z)$, thereby increasing radiative cooling efficiency at these radii. This enhanced cooling raises the gas density around $\sim r_{\rm eff}$ (Fig.~\ref{fig:profiles}, top left), leading to an increase in X-ray luminosity (Fig.~\ref{fig:lx_t_relation}). The cooled gas subsequently falls back toward the center, sustaining SMBH accretion. This coupled process establishes a galactic fountain–like circulation, in which gas is repeatedly lifted by AGN outflows and then falls back. As a consequence, inflow and outflow largely cancel in a time-averaged sense, yielding a radial velocity close to zero (Fig.~\ref{fig:profiles}, bottom left), while continuously redistributing metals throughout the inner halo.

We also found a systematic discrepancy between the mass-weighted temperature ($T_{\rm mw}$) and the \textsc{AtomDB}-derived X-ray luminosity-weighted temperature ($T_{\rm X}$). In all dE models, $T_{\rm X}$ exceeds $T_{\rm mw}$, and their rank ordering is reversed ($T_{\rm X}$: \texttt{Fiducial} $\gtrsim$ \texttt{noAGN} $>$ \texttt{noSN}; $T_{\rm mw}$: \texttt{Fiducial} $<$ \texttt{noAGN} $<$ \texttt{noSN}). This discrepancy reflects the emissivity weighting. The mass-weighted temperature includes all gas phases and is therefore sensitive to cold gas accumulation, whereas $T_{\rm X}$ is dominated by the densest X-ray–emitting gas ($\epsilon \propto \rho^{2}\Lambda$) and is largely insensitive to gas at $T \lesssim \mathrm{few}\times 10^{4}\,\mathrm{K}$.

In the \texttt{noSN} run, core depletion limits the subsequent cooling, leaving little cold gas and maintaining a relatively high $T_{\rm mw}$. In contrast, the cooling in the \texttt{Fiducial} run is stronger, driven by metal-enriched circulation, leading to significant cold gas accumulation within $r_{\rm eff}$, and thus, to a lower $T_{\rm mw}$. The ordering of $T_{\rm X}$ is determined by the distribution of X-ray–emitting gas in density–temperature space; a detailed decomposition is beyond the scope of this work.

    \section{Discussion\label{sec:discussion}}
    \subsection{Effects of kinetic jet feedback in massive halos\label{sec:disc_jets}}

    Fig. \ref{fig:lx_t_relation} shows that our fiducial CCG model lies below the
    observed best-fit $L_{\rm X}$--$T$ relation, which was extrapolated 
    beyond the ETG mass range probed by \citet{goulding16}. We note that this model does not include 
    jet feedback, whereas a jet is widely thought to play an important 
    feedback role in galaxy clusters. To explore how the inclusion of a 
    jet might shift the predicted $L_{\rm X}$ and $T$, we compared our results 
    with the recent galaxy cluster simulations presented in \citet{he26}. 
    These authors simulated the cooling flow problem in a galaxy
    cluster using the same MACER3D framework. They found that when wind and
    a jet are included, the strong shearing between them produces strong turbulence,
    which efficiently dissipates the kinetic energy of the jet and wind, and thus
    successfully solves the cooling flow problem. Three models were considered
    in that work, namely the \texttt{JetOnly}, \texttt{WindOnly}, and \texttt{JetWind}
    models. The physics included in their \texttt{JetWind} model is exactly the same
    as in the fiducial model, but includes a jet. 
    We briefly introduce the simulation of \citet{he26} in Appendix~\ref{sec:appendix_a} and refer to \citet{he26} for details. We post-processed their simulation snapshots with exactly the same synthetic
    X-ray pipeline as adopted for our models, using their snapshots and
    configuration files.
    
    The predictions of the \texttt{WindOnly} and \texttt{JetWind} presented in
    \citet{he26} are shown in Fig. \ref{fig:lx_t_relation}. The figure shows
    that their \texttt{WindOnly} model shows a similar qualitative offset in the
    $L_{\rm X}$--$T$ plane as our fiducial model, while the prediction of the
    \texttt{JetWind} model matches the observation very well. The higher
    luminosity predicted by the \texttt{JetWind} is due to its corresponding higher
    density, as shown in Fig.~\ref{fig:appendix_profiles} in Appendix~\ref{sec:appendix_a}.
    The detailed analysis presented in \citet{he26} indicates that when both wind
    and jet are included, the strong shear between them produces strong turbulence.
    The turbulence efficiently dissipates the kinetic energy of the wind and jet and heats the gas at relatively small radii of the galaxy. Consequently, the black hole accretion rate is lower than in the \texttt{WindOnly} model. The wind launched from such a low-power AGN is not able to expel the gas at relatively large radii, and thus, the density is high.

    \subsection{Systematic uncertainties and future extensions\label{sec:disc_caveats}}
    Nonthermal components, such as magnetic fields and cosmic rays (CRs), can modify
    the thermodynamic state of galactic atmospheres by contributing additional pressure
    support and changing the buoyancy and transport of feedback-heated gas.
    Their impact appears to depend strongly on the mass. At the low-mass end,
    magnetohydrodynamic (MHD) effects tend to reduce the X-ray output relative
    to purely hydrodynamic treatments \citep{su19}, and the inclusion of
    magnetic fields can lower the central gas temperature in dwarf systems
    \citep{su17}. By contrast, the influence of magnetic fields
    and CRs on the integrated X-ray luminosity in massive halos is reported to be small \citep{su19},
    and magnetic fields have little effect on the characteristic gas temperature
    in Milky Way-mass systems \citep{su17}.

    Cosmic rays introduce an additional mass dependence beyond MHD
    alone. \citet{ji20} reported that for dwarf galaxies, adding CRs to
    magnetic fields did not significantly change the gas temperature compared
    to MHD runs, whereas the CR-induced temperature reduction became increasingly
    pronounced with rising stellar mass, particularly in the CGM. At the same time, they reported that the inclusion of CRs did not
    strongly alter the gas-density profiles across galaxy mass when compared to
    the corresponding MHD models.

    Environmental processes provide a complementary mechanism, especially for dwarfs
    that are commonly satellites. Ram-pressure and tidal stripping can
    remove gas from the system \citep{mayer06}, which naturally lowers the
    emission measure and hence the X-ray luminosity; preferential removal of low-entropy
    gas can also lead to a modest increase in the characteristic temperature of the
    remaining hot phase. For massive galaxies, the sensitivity to the external gas
    supply depends on how efficiently cosmological inflows penetrate to small radii.
    In the cosmological simulations of \citet{zhu23a}, the fiducial inflow did
    not reach the central galaxy and left the gas density within one effective radius
    nearly unchanged while slightly increasing the temperature; however, an enhanced
    inflow (by a factor of $\sim 3$) substantially increased the central gas density
    and modified the temperature profile. Future simulations that include nonthermal physics and environmental
    interactions  will be important for clarifying how
    they affect the observed $L_{\rm X}$--$T$ relation.

    An additional systematic uncertainty stems from the initial redshift 
    $z\sim2$, which was adopted to set the stellar population age, but does not 
    reflect a fully cosmological setup (Section~\ref{sec:results}). The younger 
    stellar age leads to higher stellar wind and Type Ia SN rates than in 
    present-day galaxies. Type II SNe, triggered by star formation within 
    the simulation, are unaffected. For the massive galaxies, where AGN feedback 
    dominates the energy budget, this overestimate does not alter our qualitative 
    conclusions. For the dwarf ellipticals, Type II SNe dominate the 
    stellar feedback energy budget and are self-consistently regulated by 
    simulated star formation. The overestimated channels mainly contribute to 
    the mass and metal supply and not to the total heating. The enhanced 
    metal enrichment can moderately boost radiative cooling, but the net effect 
    on our conclusions is expected to be modest.

    \subsection{Comparison with previous studies}

    Our mE model overlaps with the 
    mass range studied by \citet{pellegrini18}, \citet{ciotti22}, and 
    \citet{pellegrini25}, who used 2D hydrodynamical simulations to study X-ray 
    emission from ETGs. We confirm their conclusion that AGN feedback is 
    essential for maintaining a quasi-steady state in massive systems. However, 
    whereas \citet{pellegrini18} reported that AGN feedback does not 
    significantly alter $L_{\rm X}$, our 3D simulation shows a reduction by more 
    than an order of magnitude when AGN feedback is included (compare 
    \texttt{Fiducial} and \texttt{noAGN} in Fig.~\ref{fig:lx_t_relation}). A more probable cause is our 
    GRMHD-constrained AGN wind model, together with 3D multiphase mixing that 
    might modify the cooling efficiency relative to 2D treatments. Beyond this 
    point, our setup differs from these earlier works in two additional respects. 
    We did not include CGM accretion, which \citet{pellegrini25} found to be 
    necessary to reproduce the outer X-ray surface brightness and temperature 
    profiles. Our models are also nonrotating, whereas \citet{ciotti22} showed 
    that rotation and the associated formation of cold gas disks can 
    significantly reduce $L_{\rm X}$.

    \section{Conclusion\label{sec:conclusion}}
    We used the 3D MACER simulation framework to investigate the thermodynamics of hot galactic atmospheres in dwarf ellipticals, massive ellipticals, and cluster-central galaxies. We focused on how the interplay among AGN feedback, SN feedback, and gravitational heating regulates the X-ray luminosity and temperature, and we compared our theoretical predictions with observations. To this end, we considered three models for each system: \texttt{noAGN}, \texttt{noSN}, and a \texttt{Fiducial} model that included AGN and SN feedback. Our main findings are summarized below.

    \begin{itemize}
        \item The CCG case had the deepest gravitational potential. Neither AGN winds nor SN heating alone was sufficient to reproduce the observed temperature–luminosity relation; in both cases, the predicted luminosity was too high for a given temperature. When we included AGN and SN feedback (the \texttt{Fiducial} model), their nonlinear coupling suppressed the luminosity far more efficiently than either process alone, yielding values that were lower by more than an order of magnitude than in the \texttt{noAGN} or \texttt{noSN} runs. However, in this case, the combined effect suppressed $L_{\rm X}$ well below the 
        extrapolation of the observed ETG relation.

        \item To understand this discrepancy, we analyzed the cluster simulations of \citet{he26}. In contrast to our \texttt{Fiducial} model, their model additionally included AGN jet feedback. The predicted luminosity in their \texttt{WindJet} run is consistent with observations. Physically, the inclusion of jets reduces the gas density near the black hole, thereby lowering the accretion rate. As a result, the AGN-driven winds weaken, and higher gas densities can be maintained at larger radii, which increases the X-ray luminosity.
        
        \item In intermediate-mass elliptical galaxies, SN feedback alone (i.e., the \texttt{noAGN} model) is insufficient to suppress the gas density, resulting in X-ray luminosities that are higher than observed. When AGN feedback is included, with or without SN feedback, episodic AGN outbursts drive strong winds that expel gas from the central region and produce a flatter density profile. Consequently, the X-ray luminosity is significantly reduced and becomes consistent with observations.
        
        \item In the shallow gravitational potential of dwarf ellipticals, we found (somewhat surprisingly) that the predicted X-ray luminosity in the \texttt{Fiducial} model (including SN and AGN feedback) was higher than in models with either SN or AGN feedback alone. This behavior arises from the coupled effects of SN enrichment and AGN-driven gas transport. AGN winds lift metal-enriched gas produced by SNe from the central region to intermediate radii. The resulting metal enrichment significantly enhances the radiative cooling efficiency, increasing the gas density, and thereby boosting $L_{\rm X}$. This prediction for the location of dwarf ellipticals in the $L_{\rm X}$--$T$ plane can be tested by future observations. We also found that the cooled gas subsequently fell back toward the center, establishing a galaxy-scale fountain-like circulation.
    \end{itemize}
   
    In summary, our results indicate that the $L_{\rm X}$--$T$ scaling relation emerges from distinct hydrodynamic regimes governed by the interplay among SN feedback, AGN feedback, and the gravitational potential. Future extensions of the MACER framework that incorporate additional physics (e.g., magnetic fields, cosmic-ray transport, and cosmological or environmental boundary conditions) will be essential for placing tighter constraints on the thermodynamic state of galaxies.

    \begin{acknowledgements}
        We thank the anonymous referee for the constructive comments which greatly improve the presentation of the paper. H.X., F.Y. and S.J. are supported by Natural Science Foundation of China (grants
        No. 12133008, 12192220, 12522301, 12192223, and 12361161601), China Manned Space
        Project (grants CMS-CSST-2025-A08 and CMS-CSST-2025-A10), the National Key R\&D Program of China No. 2023YFB3002502 and the National SKA Program of China
        (No. 2025SKA0130100). Numerical calculations were run on the supercomputing
        system in the Supercomputing Center of Wuhan University and the High Performance
        Computing Resource in the Core Facility for Advanced Research Computing at
        Shanghai Astronomical Observatory.
    \end{acknowledgements}
    \bibliographystyle{aa}
    \bibliography{ref}

@article{allen11,
  author        = {{Allen}, Steven W. and {Evrard}, August E. and {Mantz}, Adam B.},
  title         = {{Cosmological Parameters from Observations of Galaxy Clusters}},
  journal       = {\araa},
  year          = 2011,
  month         = sep,
  volume        = {49},
  number        = {1},
  pages         = {409-470},
  doi           = {10.1146/annurev-astro-081710-102514},
  archiveprefix = {arXiv},
  eprint        = {1103.4829},
  primaryclass  = {astro-ph.CO},
  adsurl        = {https://ui.adsabs.harvard.edu/abs/2011ARA&A..49..409A}
}

@article{arnaud99,
  author        = {{Arnaud}, Monique and {Evrard}, August E.},
  title         = {{The L\_X-T relation and intracluster gas fractions of X-ray clusters}},
  journal       = {\mnras},
  year          = 1999,
  month         = may,
  volume        = {305},
  number        = {3},
  pages         = {631-640},
  doi           = {10.1046/j.1365-8711.1999.02442.x},
  archiveprefix = {arXiv},
  eprint        = {astro-ph/9806353},
  primaryclass  = {astro-ph},
  adsurl        = {https://ui.adsabs.harvard.edu/abs/1999MNRAS.305..631A}
}

@article{athena,
  author        = {{Stone}, James M. and {Tomida}, Kengo and {White}, Christopher J. and {Felker}, Kyle G.},
  title         = {{The Athena++ Adaptive Mesh Refinement Framework: Design and Magnetohydrodynamic Solvers}},
  journal       = {\apjs},
  year          = 2020,
  month         = jul,
  volume        = {249},
  number        = {1},
  eid           = {4},
  pages         = {4},
  doi           = {10.3847/1538-4365/ab929b},
  archiveprefix = {arXiv},
  eprint        = {2005.06651},
  primaryclass  = {astro-ph.IM},
  adsurl        = {https://ui.adsabs.harvard.edu/abs/2020ApJS..249....4S}
}

@article{babyk18,
  author        = {{Babyk}, Iu. V. and {McNamara}, B.~R. and {Nulsen}, P.~E.~J. and {Hogan}, M.~T. and {Vantyghem}, A.~N. and {Russell}, H.~R. and {Pulido}, F.~A. and {Edge}, A.~C.},
  title         = {{X-Ray Scaling Relations of Early-type Galaxies}},
  journal       = {\apj},
  year          = 2018,
  month         = apr,
  volume        = {857},
  number        = {1},
  eid           = {32},
  pages         = {32},
  doi           = {10.3847/1538-4357/aab3c9},
  archiveprefix = {arXiv},
  eprint        = {1803.00020},
  primaryclass  = {astro-ph.GA},
  adsurl        = {https://ui.adsabs.harvard.edu/abs/2018ApJ...857...32B}
}

@article{bahar22,
  author        = {{Bahar}, Y. Emre and {Bulbul}, Esra and {Clerc}, Nicolas and {Ghirardini}, Vittorio and {Liu}, Ang and {Nandra}, Kirpal and {Pacaud}, Florian and {Chiu}, I.-Non and {Comparat}, Johan and {Ider-Chitham}, Jacob and et al.},
  title         = {{The eROSITA Final Equatorial-Depth Survey (eFEDS). X-ray properties and scaling relations of galaxy clusters and groups}},
  journal       = {\aap},
  year          = 2022,
  month         = may,
  volume        = {661},
  eid           = {A7},
  pages         = {A7},
  doi           = {10.1051/0004-6361/202142462},
  archiveprefix = {arXiv},
  eprint        = {2110.09534},
  primaryclass  = {astro-ph.CO},
  adsurl        = {https://ui.adsabs.harvard.edu/abs/2022A&A...661A...7B}
}

@article{balogh01,
  author        = {{Balogh}, Michael L. and {Pearce}, Frazer R. and {Bower}, Richard G. and {Kay}, Scott T.},
  title         = {{Revisiting the cosmic cooling crisis}},
  journal       = {\mnras},
  year          = 2001,
  month         = oct,
  volume        = {326},
  number        = {4},
  pages         = {1228-1234},
  doi           = {10.1111/j.1365-2966.2001.04667.x},
  archiveprefix = {arXiv},
  eprint        = {astro-ph/0104041},
  primaryclass  = {astro-ph},
  adsurl        = {https://ui.adsabs.harvard.edu/abs/2001MNRAS.326.1228B}
}

@article{blumenthal84,
  author   = {{Blumenthal}, G.~R. and {Faber}, S.~M. and {Primack}, J.~R. and {Rees}, M.~J.},
  title    = {{Formation of galaxies and large-scale structure with cold dark matter.}},
  journal  = {\nat},
  year     = 1984,
  month    = oct,
  volume   = {311},
  pages    = {517-525},
  doi      = {10.1038/311517a0},
  adsurl   = {https://ui.adsabs.harvard.edu/abs/1984Natur.311..517B}
}

@article{borgani04,
  author        = {{Borgani}, S. and {Murante}, G. and {Springel}, V. and {Diaferio}, A. and {Dolag}, K. and {Moscardini}, L. and {Tormen}, G. and {Tornatore}, L. and {Tozzi}, P.},
  title         = {{X-ray properties of galaxy clusters and groups from a cosmological hydrodynamical simulation}},
  journal       = {\mnras},
  year          = 2004,
  month         = mar,
  volume        = {348},
  number        = {3},
  pages         = {1078-1096},
  doi           = {10.1111/j.1365-2966.2004.07431.x},
  archiveprefix = {arXiv},
  eprint        = {astro-ph/0310794},
  primaryclass  = {astro-ph},
  adsurl        = {https://ui.adsabs.harvard.edu/abs/2004MNRAS.348.1078B}
}

@article{boroson11,
  author        = {{Boroson}, Bram and {Kim}, Dong-Woo and {Fabbiano}, Giuseppina},
  title         = {{Revisiting with Chandra the Scaling Relations of the X-ray Emission Components (Binaries, Nuclei, and Hot Gas) of Early-type Galaxies}},
  journal       = {\apj},
  year          = 2011,
  month         = mar,
  volume        = {729},
  number        = {1},
  eid           = {12},
  pages         = {12},
  doi           = {10.1088/0004-637X/729/1/12},
  archiveprefix = {arXiv},
  eprint        = {1011.2529},
  primaryclass  = {astro-ph.HE},
  adsurl        = {https://ui.adsabs.harvard.edu/abs/2011ApJ...729...12B}
}

@article{bourne17,
  author        = {{Bourne}, Martin A. and {Sijacki}, Debora},
  title         = {{AGN jet feedback on a moving mesh: cocoon inflation, gas flows and turbulence}},
  journal       = {\mnras},
  year          = 2017,
  month         = dec,
  volume        = {472},
  number        = {4},
  pages         = {4707-4735},
  doi           = {10.1093/mnras/stx2269},
  archiveprefix = {arXiv},
  eprint        = {1705.07900},
  primaryclass  = {astro-ph.GA},
  adsurl        = {https://ui.adsabs.harvard.edu/abs/2017MNRAS.472.4707B}
}

@article{bryan98,
  author        = {{Bryan}, Greg L. and {Norman}, Michael L.},
  title         = {{Statistical Properties of X-Ray Clusters: Analytic and Numerical Comparisons}},
  journal       = {\apj},
  year          = 1998,
  month         = mar,
  volume        = {495},
  number        = {1},
  pages         = {80-99},
  doi           = {10.1086/305262},
  archiveprefix = {arXiv},
  eprint        = {astro-ph/9710107},
  primaryclass  = {astro-ph},
  adsurl        = {https://ui.adsabs.harvard.edu/abs/1998ApJ...495...80B}
}

@article{churazov01,
  author        = {{Churazov}, E. and {Br{\"u}ggen}, M. and {Kaiser}, C.~R. and {B{\"o}hringer}, H. and {Forman}, W.},
  title         = {{Evolution of Buoyant Bubbles in M87}},
  journal       = {\apj},
  year          = 2001,
  month         = jun,
  volume        = {554},
  number        = {1},
  pages         = {261-273},
  doi           = {10.1086/321357},
  archiveprefix = {arXiv},
  eprint        = {astro-ph/0008215},
  primaryclass  = {astro-ph},
  adsurl        = {https://ui.adsabs.harvard.edu/abs/2001ApJ...554..261C}
}

@article{ciotti09,
  author        = {{Ciotti}, Luca and {Ostriker}, Jeremiah P. and {Proga}, Daniel},
  title         = {{Feedback from Central Black Holes in Elliptical Galaxies. I. Models with Either Radiative or Mechanical Feedback but not Both}},
  journal       = {\apj},
  year          = 2009,
  month         = jul,
  volume        = {699},
  number        = {1},
  pages         = {89-104},
  doi           = {10.1088/0004-637X/699/1/89},
  archiveprefix = {arXiv},
  eprint        = {0901.1089},
  primaryclass  = {astro-ph.GA},
  adsurl        = {https://ui.adsabs.harvard.edu/abs/2009ApJ...699...89C}
}

@article{ciotti10,
  author        = {{Ciotti}, Luca and {Ostriker}, Jeremiah P. and {Proga}, Daniel},
  title         = {{Feedback from Central Black Holes in Elliptical Galaxies. III. Models with Both Radiative and Mechanical Feedback}},
  journal       = {\apj},
  year          = 2010,
  month         = jul,
  volume        = {717},
  number        = {2},
  pages         = {708-723},
  doi           = {10.1088/0004-637X/717/2/708},
  archiveprefix = {arXiv},
  eprint        = {1003.0578},
  primaryclass  = {astro-ph.CO},
  adsurl        = {https://ui.adsabs.harvard.edu/abs/2010ApJ...717..708C}
}

@inproceedings{ciotti12,
  author        = {{Ciotti}, Luca and {Ostriker}, Jeremiah P.},
  title         = {{AGN Feedback in Elliptical Galaxies: Numerical Simulations}},
  booktitle     = {Astrophysics and Space Science Library},
  year          = 2012,
  editor        = {{Kim}, Dong-Woo and {Pellegrini}, Silvia},
  series        = {Astrophysics and Space Science Library},
  volume        = {378},
  month         = jan,
  pages         = {83},
  doi           = {10.1007/978-1-4614-0580-1_4},
  archiveprefix = {arXiv},
  eprint        = {1104.2238},
  primaryclass  = {astro-ph.GA},
  adsurl        = {https://ui.adsabs.harvard.edu/abs/2012ASSL..378...83C}
}

@article{ciotti17,
  author        = {{Ciotti}, Luca and {Pellegrini}, Silvia and {Negri}, Andrea and {Ostriker}, Jeremiah P.},
  title         = {{The Effect of the AGN Feedback on the Interstellar Medium of Early-Type Galaxies:2D Hydrodynamical Simulations of the Low-Rotation Case.}},
  journal       = {\apj},
  year          = 2017,
  month         = jan,
  volume        = {835},
  number        = {1},
  eid           = {15},
  pages         = {15},
  doi           = {10.3847/1538-4357/835/1/15},
  archiveprefix = {arXiv},
  eprint        = {1608.03403},
  primaryclass  = {astro-ph.GA},
  adsurl        = {https://ui.adsabs.harvard.edu/abs/2017ApJ...835...15C}
}

@ARTICLE{ciotti22,
       author = {{Ciotti}, Luca and {Ostriker}, Jeremiah P. and {Gan}, Zhaoming and {Jiang}, Brian Xing and {Pellegrini}, Silvia and {Caravita}, Caterina and {Mancino}, Antonio},
        title = "{A Parameter Space Exploration of High-resolution Numerically Evolved Early Type Galaxies Including AGN Feedback and Accurate Dynamical Treatment of Stellar Orbits}",
      journal = {\apj},
         year = 2022,
        month = jul,
       volume = {933},
       number = {2},
          eid = {154},
        pages = {154},
          doi = {10.3847/1538-4357/ac70c7},
archivePrefix = {arXiv},
       eprint = {2201.03909},
 primaryClass = {astro-ph.GA},
       adsurl = {https://ui.adsabs.harvard.edu/abs/2022ApJ...933..154C}
}

@article{cloudy,
  author        = {{Ferland}, G.~J. and {Chatzikos}, M. and {Guzm{\'a}n}, F. and {Lykins}, M.~L. and {van Hoof}, P.~A.~M. and {Williams}, R.~J.~R. and {Abel}, N.~P. and {Badnell}, N.~R. and {Keenan}, F.~P. and {Porter}, R.~L. and {Stancil}, P.~C.},
  title         = {{The 2017 Release Cloudy}},
  journal       = {\rmxaa},
  year          = 2017,
  month         = oct,
  volume        = {53},
  pages         = {385-438},
  doi           = {10.48550/arXiv.1705.10877},
  archiveprefix = {arXiv},
  eprint        = {1705.10877},
  primaryclass  = {astro-ph.GA},
  adsurl        = {https://ui.adsabs.harvard.edu/abs/2017RMxAA..53..385F}
}

@article{di23,
  author        = {{Di}, Yihuan and {Li}, Yuan and {Yuan}, Feng and {Shi}, Fangzheng and {Caradonna}, Mirielle},
  title         = {{Black hole feeding and feedback in a compact galaxy}},
  journal       = {\mnras},
  year          = 2023,
  month         = aug,
  volume        = {523},
  number        = {2},
  pages         = {1641-1647},
  doi           = {10.1093/mnras/stad1529},
  archiveprefix = {arXiv},
  eprint        = {2305.11415},
  primaryclass  = {astro-ph.GA},
  adsurl        = {https://ui.adsabs.harvard.edu/abs/2023MNRAS.523.1641D}
}

@article{evrard90,
  author   = {{Evrard}, August E.},
  title    = {{Formation and Evolution of X-Ray Clusters: A Hydrodynamic Simulation of the Intracluster Medium}},
  journal  = {\apj},
  year     = 1990,
  month    = nov,
  volume   = {363},
  pages    = {349},
  doi      = {10.1086/169350},
  adsurl   = {https://ui.adsabs.harvard.edu/abs/1990ApJ...363..349E}
}

@article{evrard91,
  author   = {{Evrard}, A.~E. and {Henry}, J.~P.},
  title    = {{Expectations for X-Ray Cluster Observations by the ROSAT Satellite}},
  journal  = {\apj},
  year     = 1991,
  month    = dec,
  volume   = {383},
  pages    = {95},
  doi      = {10.1086/170767},
  adsurl   = {https://ui.adsabs.harvard.edu/abs/1991ApJ...383...95E}
}

@article{faber76,
  author   = {{Faber}, S.~M. and {Jackson}, R.~E.},
  title    = {{Velocity dispersions and mass-to-light ratios for elliptical galaxies.}},
  journal  = {\apj},
  year     = 1976,
  month    = mar,
  volume   = {204},
  pages    = {668-683},
  doi      = {10.1086/154215},
  adsurl   = {https://ui.adsabs.harvard.edu/abs/1976ApJ...204..668F}
}

@article{fabian12,
  author        = {{Fabian}, A.~C.},
  title         = {{Observational Evidence of Active Galactic Nuclei Feedback}},
  journal       = {\araa},
  year          = 2012,
  month         = sep,
  volume        = {50},
  pages         = {455-489},
  doi           = {10.1146/annurev-astro-081811-125521},
  archiveprefix = {arXiv},
  eprint        = {1204.4114},
  primaryclass  = {astro-ph.CO},
  adsurl        = {https://ui.adsabs.harvard.edu/abs/2012ARA&A..50..455F}
}

@article{fabian94,
  author  = {{Fabian}, A.~C.},
  title   = {{Cooling Flows in Clusters of Galaxies}},
  journal = {\araa},
  year    = 1994,
  month   = jan,
  volume  = {32},
  pages   = {277-318},
  doi     = {10.1146/annurev.aa.32.090194.001425},
  adsurl  = {https://ui.adsabs.harvard.edu/abs/1994ARA&A..32..277F}
}

@article{foster12,
  author        = {{Foster}, A.~R. and {Ji}, L. and {Smith}, R.~K. and {Brickhouse}, N.~S.},
  title         = {{Updated Atomic Data and Calculations for X-Ray Spectroscopy}},
  journal       = {\apj},
  year          = 2012,
  month         = sep,
  volume        = {756},
  number        = {2},
  eid           = {128},
  pages         = {128},
  doi           = {10.1088/0004-637X/756/2/128},
  archiveprefix = {arXiv},
  eprint        = {1207.0576},
  primaryclass  = {astro-ph.HE},
  adsurl        = {https://ui.adsabs.harvard.edu/abs/2012ApJ...756..128F}
}

@article{gan14,
  author        = {{Gan}, Zhaoming and {Yuan}, Feng and {Ostriker}, Jeremiah P. and {Ciotti}, Luca and {Novak}, Gregory S.},
  title         = {{Active Galactic Nucleus Feedback in an Isolated Elliptical Galaxy: The Effect of Strong Radiative Feedback in the Kinetic Mode}},
  journal       = {\apj},
  year          = 2014,
  month         = jul,
  volume        = {789},
  number        = {2},
  eid           = {150},
  pages         = {150},
  doi           = {10.1088/0004-637X/789/2/150},
  archiveprefix = {arXiv},
  eprint        = {1403.0670},
  primaryclass  = {astro-ph.GA},
  adsurl        = {https://ui.adsabs.harvard.edu/abs/2014ApJ...789..150G}
}

@article{giodini13,
  author        = {{Giodini}, S. and {Lovisari}, L. and {Pointecouteau}, E. and {Ettori}, S. and {Reiprich}, T.~H. and {Hoekstra}, H.},
  title         = {{Scaling Relations for Galaxy Clusters: Properties and Evolution}},
  journal       = {\ssr},
  year          = 2013,
  month         = aug,
  volume        = {177},
  number        = {1-4},
  pages         = {247-282},
  doi           = {10.1007/s11214-013-9994-5},
  archiveprefix = {arXiv},
  eprint        = {1305.3286},
  primaryclass  = {astro-ph.CO},
  adsurl        = {https://ui.adsabs.harvard.edu/abs/2013SSRv..177..247G}
}

@article{gofford15,
  author        = {{Gofford}, J. and {Reeves}, J.~N. and {McLaughlin}, D.~E. and {Braito}, V. and {Turner}, T.~J. and {Tombesi}, F. and {Cappi}, M.},
  title         = {{The Suzaku view of highly ionized outflows in AGN - II. Location, energetics and scalings with bolometric luminosity}},
  journal       = {\mnras},
  year          = 2015,
  month         = aug,
  volume        = {451},
  number        = {4},
  pages         = {4169-4182},
  doi           = {10.1093/mnras/stv1207},
  archiveprefix = {arXiv},
  eprint        = {1506.00614},
  primaryclass  = {astro-ph.HE},
  adsurl        = {https://ui.adsabs.harvard.edu/abs/2015MNRAS.451.4169G}
}

@article{goulding16,
  author        = {{Goulding}, Andy D. and {Greene}, Jenny E. and {Ma}, Chung-Pei and {Veale}, Melanie and {Bogdan}, Akos and {Nyland}, Kristina and {Blakeslee}, John P. and {McConnell}, Nicholas J. and {Thomas}, Jens},
  title         = {{The MASSIVE Survey. IV. The X-ray Halos of the Most Massive Early-type Galaxies in the Nearby Universe}},
  journal       = {\apj},
  year          = 2016,
  month         = aug,
  volume        = {826},
  number        = {2},
  eid           = {167},
  pages         = {167},
  doi           = {10.3847/0004-637X/826/2/167},
  archiveprefix = {arXiv},
  eprint        = {1604.01764},
  primaryclass  = {astro-ph.GA},
  adsurl        = {https://ui.adsabs.harvard.edu/abs/2016ApJ...826..167G}
}

@article{harrison18,
  author        = {{Harrison}, C.~M. and {Costa}, T. and {Tadhunter}, C.~N. and {Fl{\"u}tsch}, A. and {Kakkad}, D. and {Perna}, M. and {Vietri}, G.},
  title         = {{AGN outflows and feedback twenty years on}},
  journal       = {Nature Astronomy},
  year          = 2018,
  month         = feb,
  volume        = {2},
  pages         = {198-205},
  doi           = {10.1038/s41550-018-0403-6},
  archiveprefix = {arXiv},
  eprint        = {1802.10306},
  primaryclass  = {astro-ph.GA},
  adsurl        = {https://ui.adsabs.harvard.edu/abs/2018NatAs...2..198H}
}

@ARTICLE{he26,
       author = {{He}, Aoyun and {Yuan}, Feng and {Ji}, Suoqing and {Guo}, Minhang and {Li}, Yuan and {Xu}, Haiguang and {Sun}, Ming and {Xia}, Haojie and {Zhao}, Yuanyuan},
        title = "{Solving the cooling flow problem with combined jet-wind AGN feedback}",
      journal = {Science Advances},
         year = 2026,
        month = jul,
       volume = {12},
        pages = {4.6394},
          doi = {10.1126/sciadv.aed6394},
       adsurl = {https://ui.adsabs.harvard.edu/abs/2026SciA...12d6394H}
}

@article{hopkins16,
  author        = {{Hopkins}, Philip F. and {Torrey}, Paul and {Faucher-Gigu{\`e}re}, Claude-Andr{\'e} and {Quataert}, Eliot and {Murray}, Norman},
  title         = {{Stellar and quasar feedback in concert: effects on AGN accretion, obscuration, and outflows}},
  journal       = {\mnras},
  year          = 2016,
  month         = may,
  volume        = {458},
  number        = {1},
  pages         = {816-831},
  doi           = {10.1093/mnras/stw289},
  archiveprefix = {arXiv},
  eprint        = {1504.05209},
  primaryclass  = {astro-ph.GA},
  adsurl        = {https://ui.adsabs.harvard.edu/abs/2016MNRAS.458..816H}
}

@article{hopkins18,
  author        = {{Hopkins}, Philip F. and {Wetzel}, Andrew and {Kere{\v{s}}}, Du{\v{s}}an and {Faucher-Gigu{\`e}re}, Claude-Andr{\'e} and {Quataert}, Eliot and {Boylan-Kolchin}, Michael and {Murray}, Norman and {Hayward}, Christopher C. and {Garrison-Kimmel}, Shea and {Hummels}, Cameron and {Feldmann}, Robert and {Torrey}, Paul and {Ma}, Xiangcheng and {Angl{\'e}s-Alc{\'a}zar}, Daniel and {Su}, Kung-Yi and {Orr}, Matthew and {Schmitz}, Denise and {Escala}, Ivanna and {Sanderson}, Robyn and {Grudi{\'c}}, Michael Y. and {Hafen}, Zachary and {Kim}, Ji-Hoon and {Fitts}, Alex and {Bullock}, James S. and {Wheeler}, Coral and {Chan}, T.~K. and {Elbert}, Oliver D. and {Narayanan}, Desika},
  title         = {{FIRE-2 simulations: physics versus numerics in galaxy formation}},
  journal       = {MNRAS},
  year          = 2018,
  month         = oct,
  volume        = {480},
  number        = {1},
  pages         = {800-863},
  doi           = {10.1093/mnras/sty1690},
  archiveprefix = {arXiv},
  eprint        = {1702.06148},
  primaryclass  = {astro-ph.GA},
  adsurl        = {https://ui.adsabs.harvard.edu/abs/2018MNRAS.480..800H}
}

@article{jaffe83,
  author   = {{Jaffe}, W.},
  title    = {{A simple model for the distribution of light in spherical galaxies.}},
  journal  = {\mnras},
  year     = 1983,
  month    = mar,
  volume   = {202},
  pages    = {995-999},
  doi      = {10.1093/mnras/202.4.995},
  adsurl   = {https://ui.adsabs.harvard.edu/abs/1983MNRAS.202..995J}
}

@article{ji20,
  author        = {{Ji}, Suoqing and {Chan}, T.~K. and {Hummels}, Cameron B. and {Hopkins}, Philip F. and {Stern}, Jonathan and {Kere{\v{s}}}, Du{\v{s}}an and {Quataert}, Eliot and {Faucher-Gigu{\`e}re}, Claude-Andr{\'e} and {Murray}, Norman},
  title         = {{Properties of the circumgalactic medium in cosmic ray-dominated galaxy haloes}},
  journal       = {\mnras},
  year          = 2020,
  month         = aug,
  volume        = {496},
  number        = {4},
  pages         = {4221-4238},
  doi           = {10.1093/mnras/staa1849},
  archiveprefix = {arXiv},
  eprint        = {1909.00003},
  primaryclass  = {astro-ph.GA},
  adsurl        = {https://ui.adsabs.harvard.edu/abs/2020MNRAS.496.4221J}
}

@article{jiang19,
  author        = {{Jiang}, Yan-Fei and {Blaes}, Omer and {Stone}, James M. and {Davis}, Shane W.},
  title         = {{Global Radiation Magnetohydrodynamic Simulations of sub-Eddington Accretion Disks around Supermassive Black Holes}},
  journal       = {\apj},
  year          = 2019,
  month         = nov,
  volume        = {885},
  number        = {2},
  eid           = {144},
  pages         = {144},
  doi           = {10.3847/1538-4357/ab4a00},
  archiveprefix = {arXiv},
  eprint        = {1904.01674},
  primaryclass  = {astro-ph.HE},
  adsurl        = {https://ui.adsabs.harvard.edu/abs/2019ApJ...885..144J}
}

@article{kaiser86,
  author   = {{Kaiser}, N.},
  title    = {{Evolution and clustering of rich clusters.}},
  journal  = {\mnras},
  year     = 1986,
  month    = sep,
  volume   = {222},
  pages    = {323-345},
  doi      = {10.1093/mnras/222.2.323},
  adsurl   = {https://ui.adsabs.harvard.edu/abs/1986MNRAS.222..323K}
}

@article{khalil24,
  author        = {{Khalil}, H. and {Finoguenov}, A. and {Tempel}, E. and {Mamon}, G.~A.},
  title         = {{AXES-2MRS: A new all-sky catalogue of extended X-ray galaxy groups}},
  journal       = {\aap},
  year          = 2024,
  month         = oct,
  volume        = {690},
  eid           = {A212},
  pages         = {A212},
  doi           = {10.1051/0004-6361/202450060},
  archiveprefix = {arXiv},
  eprint        = {2403.17061},
  primaryclass  = {astro-ph.GA},
  adsurl        = {https://ui.adsabs.harvard.edu/abs/2024A&A...690A.212K}
}

@article{kim15,
  author        = {{Kim}, Dong-Woo and {Fabbiano}, Giuseppina},
  title         = {{X-Ray Scaling Relations of 'Core' and 'Coreless' E and S0 Galaxies}},
  journal       = {\apj},
  year          = 2015,
  month         = oct,
  volume        = {812},
  number        = {2},
  eid           = {127},
  pages         = {127},
  doi           = {10.1088/0004-637X/812/2/127},
  archiveprefix = {arXiv},
  eprint        = {1504.00899},
  primaryclass  = {astro-ph.GA},
  adsurl        = {https://ui.adsabs.harvard.edu/abs/2015ApJ...812..127K}
}

@article{kormendy13,
  author        = {{Kormendy}, John and {Ho}, Luis C.},
  title         = {{Coevolution (Or Not) of Supermassive Black Holes and Host Galaxies}},
  journal       = {\araa},
  year          = 2013,
  month         = aug,
  volume        = {51},
  number        = {1},
  pages         = {511-653},
  doi           = {10.1146/annurev-astro-082708-101811},
  archiveprefix = {arXiv},
  eprint        = {1304.7762},
  primaryclass  = {astro-ph.CO},
  adsurl        = {https://ui.adsabs.harvard.edu/abs/2013ARA&A..51..511K}
}

@article{kravtsov12,
  author        = {{Kravtsov}, Andrey V. and {Borgani}, Stefano},
  title         = {{Formation of Galaxy Clusters}},
  journal       = {\araa},
  year          = 2012,
  month         = sep,
  volume        = {50},
  pages         = {353-409},
  doi           = {10.1146/annurev-astro-081811-125502},
  archiveprefix = {arXiv},
  eprint        = {1205.5556},
  primaryclass  = {astro-ph.CO},
  adsurl        = {https://ui.adsabs.harvard.edu/abs/2012ARA&A..50..353K}
}

@article{lovisari15,
  author        = {{Lovisari}, L. and {Reiprich}, T.~H. and {Schellenberger}, G.},
  title         = {{Scaling properties of a complete X-ray selected galaxy group sample}},
  journal       = {\aap},
  year          = 2015,
  month         = jan,
  volume        = {573},
  eid           = {A118},
  pages         = {A118},
  doi           = {10.1051/0004-6361/201423954},
  archiveprefix = {arXiv},
  eprint        = {1409.3845},
  primaryclass  = {astro-ph.CO},
  adsurl        = {https://ui.adsabs.harvard.edu/abs/2015A&A...573A.118L}
}

@article{markevitch98,
  author        = {{Markevitch}, Maxim},
  title         = {{The L$_{X}$-T Relation and Temperature Function for Nearby Clusters Revisited}},
  journal       = {\apj},
  year          = 1998,
  month         = sep,
  volume        = {504},
  number        = {1},
  pages         = {27-34},
  doi           = {10.1086/306080},
  archiveprefix = {arXiv},
  eprint        = {astro-ph/9802059},
  primaryclass  = {astro-ph},
  adsurl        = {https://ui.adsabs.harvard.edu/abs/1998ApJ...504...27M}
}

@article{mathews03,
  author        = {{Mathews}, William G. and {Brighenti}, Fabrizio},
  title         = {{Hot Gas in and around Elliptical Galaxies}},
  journal       = {\araa},
  year          = 2003,
  month         = jan,
  volume        = {41},
  pages         = {191-239},
  doi           = {10.1146/annurev.astro.41.090401.094542},
  archiveprefix = {arXiv},
  eprint        = {astro-ph/0309553},
  primaryclass  = {astro-ph},
  adsurl        = {https://ui.adsabs.harvard.edu/abs/2003ARA&A..41..191M}
}

@article{mathews06,
  author        = {{Mathews}, William G. and {Faltenbacher}, Andreas and {Brighenti}, Fabrizio},
  title         = {{Heating Cooling Flows with Weak Shock Waves}},
  journal       = {\apj},
  year          = 2006,
  month         = feb,
  volume        = {638},
  number        = {2},
  pages         = {659-667},
  doi           = {10.1086/499119},
  archiveprefix = {arXiv},
  eprint        = {astro-ph/0511151},
  primaryclass  = {astro-ph},
  adsurl        = {https://ui.adsabs.harvard.edu/abs/2006ApJ...638..659M}
}

@article{maughan12,
  author        = {{Maughan}, B.~J. and {Giles}, P.~A. and {Randall}, S.~W. and {Jones}, C. and {Forman}, W.~R.},
  title         = {{Self-similar scaling and evolution in the galaxy cluster X-ray luminosity-temperature relation}},
  journal       = {\mnras},
  year          = 2012,
  month         = apr,
  volume        = {421},
  number        = {2},
  pages         = {1583-1602},
  doi           = {10.1111/j.1365-2966.2012.20419.x},
  archiveprefix = {arXiv},
  eprint        = {1108.1200},
  primaryclass  = {astro-ph.CO},
  adsurl        = {https://ui.adsabs.harvard.edu/abs/2012MNRAS.421.1583M}
}

@article{mayer06,
  author        = {{Mayer}, Lucio and {Mastropietro}, Chiara and {Wadsley}, James and {Stadel}, Joachim and {Moore}, Ben},
  title         = {{Simultaneous ram pressure and tidal stripping; how dwarf spheroidals lost their gas}},
  journal       = {\mnras},
  year          = 2006,
  month         = jul,
  volume        = {369},
  number        = {3},
  pages         = {1021-1038},
  doi           = {10.1111/j.1365-2966.2006.10403.x},
  archiveprefix = {arXiv},
  eprint        = {astro-ph/0504277},
  primaryclass  = {astro-ph},
  adsurl        = {https://ui.adsabs.harvard.edu/abs/2006MNRAS.369.1021M}
}

@article{mazzotta04,
  author        = {{Mazzotta}, P. and {Rasia}, E. and {Moscardini}, L. and {Tormen}, G.},
  title         = {{Comparing the temperatures of galaxy clusters from hydrodynamical N-body simulations to Chandra and XMM-Newton observations}},
  journal       = {\mnras},
  year          = 2004,
  month         = oct,
  volume        = {354},
  number        = {1},
  pages         = {10-24},
  doi           = {10.1111/j.1365-2966.2004.08167.x},
  archiveprefix = {arXiv},
  eprint        = {astro-ph/0404425},
  primaryclass  = {astro-ph},
  adsurl        = {https://ui.adsabs.harvard.edu/abs/2004MNRAS.354...10M}
}

@inbook{mc06,
  author    = {{McClintock}, Jeffrey E. and {Remillard}, Ronald A.},
  title     = {{Black hole binaries}},
  booktitle = {Compact stellar X-ray sources},
  year      = 2006,
  volume    = {39},
  pages     = {157-213},
  adsurl    = {https://ui.adsabs.harvard.edu/abs/2006csxs.book..157M}
}

@article{mcnamara07,
  author        = {{McNamara}, B.~R. and {Nulsen}, P.~E.~J.},
  title         = {{Heating Hot Atmospheres with Active Galactic Nuclei}},
  journal       = {\araa},
  year          = 2007,
  month         = sep,
  volume        = {45},
  number        = {1},
  pages         = {117-175},
  doi           = {10.1146/annurev.astro.45.051806.110625},
  archiveprefix = {arXiv},
  eprint        = {0709.2152},
  primaryclass  = {astro-ph},
  adsurl        = {https://ui.adsabs.harvard.edu/abs/2007ARA&A..45..117M}
}

@article{mukherjee16,
  author        = {{Mukherjee}, Dipanjan and {Bicknell}, Geoffrey V. and {Sutherland}, Ralph and {Wagner}, Alex},
  title         = {{Relativistic jet feedback in high-redshift galaxies - I. Dynamics}},
  journal       = {\mnras},
  year          = 2016,
  month         = sep,
  volume        = {461},
  number        = {1},
  pages         = {967-983},
  doi           = {10.1093/mnras/stw1368},
  archiveprefix = {arXiv},
  eprint        = {1606.01143},
  primaryclass  = {astro-ph.HE},
  adsurl        = {https://ui.adsabs.harvard.edu/abs/2016MNRAS.461..967M}
}

@article{navarro95,
  author        = {{Navarro}, Julio F. and {Frenk}, Carlos S. and {White}, Simon D.~M.},
  title         = {{Simulations of X-ray clusters}},
  journal       = {\mnras},
  year          = 1995,
  month         = aug,
  volume        = {275},
  number        = {3},
  pages         = {720-740},
  doi           = {10.1093/mnras/275.3.720},
  archiveprefix = {arXiv},
  eprint        = {astro-ph/9408069},
  primaryclass  = {astro-ph},
  adsurl        = {https://ui.adsabs.harvard.edu/abs/1995MNRAS.275..720N}
}

@article{navarro97,
  author        = {{Navarro}, Julio F. and {Frenk}, Carlos S. and {White}, Simon D.~M.},
  title         = {{A Universal Density Profile from Hierarchical Clustering}},
  journal       = {\apj},
  year          = 1997,
  month         = dec,
  volume        = {490},
  number        = {2},
  pages         = {493-508},
  doi           = {10.1086/304888},
  archiveprefix = {arXiv},
  eprint        = {astro-ph/9611107},
  primaryclass  = {astro-ph},
  adsurl        = {https://ui.adsabs.harvard.edu/abs/1997ApJ...490..493N}
}

@article{negri17,
  author        = {{Negri}, A. and {Volonteri}, M.},
  title         = {{Black hole feeding and feedback: the physics inside the `sub-grid'}},
  journal       = {\mnras},
  year          = 2017,
  month         = may,
  volume        = {467},
  number        = {3},
  pages         = {3475-3492},
  doi           = {10.1093/mnras/stx362},
  archiveprefix = {arXiv},
  eprint        = {1610.04753},
  primaryclass  = {astro-ph.GA},
  adsurl        = {https://ui.adsabs.harvard.edu/abs/2017MNRAS.467.3475N}
}

@article{nomoto13,
  author  = {{Nomoto}, Ken'ichi and {Kobayashi}, Chiaki and {Tominaga}, Nozomu},
  title   = {{Nucleosynthesis in Stars and the Chemical Enrichment of Galaxies}},
  journal = {ARA\&A},
  year    = 2013,
  month   = aug,
  volume  = {51},
  number  = {1},
  pages   = {457-509},
  doi     = {10.1146/annurev-astro-082812-140956},
  adsurl  = {https://ui.adsabs.harvard.edu/abs/2013ARA&A..51..457N}
}

@article{ostriker10,
  author        = {{Ostriker}, Jeremiah P. and {Choi}, Ena and {Ciotti}, Luca and {Novak}, Gregory S. and {Proga}, Daniel},
  title         = {{Momentum Driving: Which Physical Processes Dominate Active Galactic Nucleus Feedback?}},
  journal       = {\apj},
  year          = 2010,
  month         = oct,
  volume        = {722},
  number        = {1},
  pages         = {642-652},
  doi           = {10.1088/0004-637X/722/1/642},
  archiveprefix = {arXiv},
  eprint        = {1004.2923},
  primaryclass  = {astro-ph.GA},
  adsurl        = {https://ui.adsabs.harvard.edu/abs/2010ApJ...722..642O}
}

@ARTICLE{pellegrini18,
       author = {{Pellegrini}, Silvia and {Ciotti}, Luca and {Negri}, Andrea and {Ostriker}, Jeremiah P.},
        title = "{Active Galactic Nuclei Feedback and the Origin and Fate of the Hot Gas in Early-type Galaxies}",
      journal = {\apj},
         year = 2018,
        month = apr,
       volume = {856},
       number = {2},
          eid = {115},
        pages = {115},
          doi = {10.3847/1538-4357/aaae07},
archivePrefix = {arXiv},
       eprint = {1802.02005},
 primaryClass = {astro-ph.GA},
       adsurl = {https://ui.adsabs.harvard.edu/abs/2018ApJ...856..115P}
}

@article{pellegrini25,
  author        = {{Pellegrini}, Silvia and {Ciotti}, Luca and {Gan}, Zhaoming and {Kim}, Dong-Woo and {Ostriker}, Jeremiah P.},
  title         = {{X-Ray Halos of Early-type Galaxies with Active Galactic Nucleus Feedback and Accretion from a Circumgalactic Medium: Models and Observations}},
  journal       = {\apj},
  year          = 2025,
  month         = sep,
  volume        = {990},
  number        = {2},
  eid           = {194},
  pages         = {194},
  doi           = {10.3847/1538-4357/adf5c1},
  archiveprefix = {arXiv},
  eprint        = {2508.03536},
  primaryclass  = {astro-ph.GA},
  adsurl        = {https://ui.adsabs.harvard.edu/abs/2025ApJ...990..194P}
}

@article{plummer11,
  author  = {{Plummer}, H.~C.},
  title   = {{On the problem of distribution in globular star clusters}},
  journal = {\mnras},
  year    = 1911,
  month   = mar,
  volume  = {71},
  pages   = {460-470},
  doi     = {10.1093/mnras/71.5.460},
  adsurl  = {https://ui.adsabs.harvard.edu/abs/1911MNRAS..71..460P}
}

@article{pratt09,
  author        = {{Pratt}, G.~W. and {Croston}, J.~H. and {Arnaud}, M. and {B{\"o}hringer}, H.},
  title         = {{Galaxy cluster X-ray luminosity scaling relations from a representative local sample (REXCESS)}},
  journal       = {\aap},
  year          = 2009,
  month         = may,
  volume        = {498},
  number        = {2},
  pages         = {361-378},
  doi           = {10.1051/0004-6361/200810994},
  archiveprefix = {arXiv},
  eprint        = {0809.3784},
  primaryclass  = {astro-ph},
  adsurl        = {https://ui.adsabs.harvard.edu/abs/2009A&A...498..361P}
}

@article{randall11,
  author        = {{Randall}, S.~W. and {Forman}, W.~R. and {Giacintucci}, S. and {Nulsen}, P.~E.~J. and {Sun}, M. and {Jones}, C. and {Churazov}, E. and {David}, L.~P. and {Kraft}, R. and {Donahue}, M. and et al.},
  title         = {{Shocks and Cavities from Multiple Outbursts in the Galaxy Group NGC 5813: A Window to Active Galactic Nucleus Feedback}},
  journal       = {\apj},
  year          = 2011,
  month         = jan,
  volume        = {726},
  number        = {2},
  eid           = {86},
  pages         = {86},
  doi           = {10.1088/0004-637X/726/2/86},
  archiveprefix = {arXiv},
  eprint        = {1006.4379},
  primaryclass  = {astro-ph.CO},
  adsurl        = {https://ui.adsabs.harvard.edu/abs/2011ApJ...726...86R}
}

@article{randall15,
  author        = {{Randall}, S.~W. and {Nulsen}, P.~E.~J. and {Jones}, C. and {Forman}, W.~R. and {Bulbul}, E. and {Clarke}, T.~E. and {Kraft}, R. and {Blanton}, E.~L. and {David}, L. and {Werner}, N. and et al.},
  title         = {{A Very Deep Chandra Observation of the Galaxy Group NGC 5813: AGN Shocks, Feedback, and Outburst History}},
  journal       = {\apj},
  year          = 2015,
  month         = jun,
  volume        = {805},
  number        = {2},
  eid           = {112},
  pages         = {112},
  doi           = {10.1088/0004-637X/805/2/112},
  archiveprefix = {arXiv},
  eprint        = {1503.08205},
  primaryclass  = {astro-ph.HE},
  adsurl        = {https://ui.adsabs.harvard.edu/abs/2015ApJ...805..112R}
}

@article{rees77,
  author   = {{Rees}, M.~J. and {Ostriker}, J.~P.},
  title    = {{Cooling, dynamics and fragmentation of massive gas clouds: clues to the masses and radii of galaxies and clusters.}},
  journal  = {\mnras},
  year     = 1977,
  month    = jun,
  volume   = {179},
  pages    = {541-559},
  doi      = {10.1093/mnras/179.4.541},
  adsurl   = {https://ui.adsabs.harvard.edu/abs/1977MNRAS.179..541R}
}

@inproceedings{sarazin86,
  author    = {{Sarazin}, Craig L.},
  title     = {{A Simple Cooling Flow Model for X-ray Coronae Around Elliptical Galaxies}},
  booktitle = {Gaseous halos of Galaxies},
  year      = 1986,
  editor    = {{Bregman}, J.~N. and {Lockman}, F.~J. and {Cox}, D.},
  month     = jan,
  pages     = {223},
  adsurl    = {https://ui.adsabs.harvard.edu/abs/1986ghg..conf..223S}
}

@article{silk77,
  author   = {{Silk}, J.},
  title    = {{On the fragmentation of cosmic gas clouds. I. The formation of galaxies and the first generation of stars.}},
  journal  = {\apj},
  year     = 1977,
  month    = feb,
  volume   = {211},
  pages    = {638-648},
  doi      = {10.1086/154972},
  adsurl   = {https://ui.adsabs.harvard.edu/abs/1977ApJ...211..638S}
}

@article{su15,
  author        = {{Su}, Yuanyuan and {Irwin}, Jimmy A. and {White}, III, Raymond E. and {Cooper}, Michael C.},
  title         = {{The Scatter in the Hot Gas Content of Early-type Galaxies}},
  journal       = {\apj},
  year          = 2015,
  month         = jun,
  volume        = {806},
  number        = {2},
  eid           = {156},
  pages         = {156},
  doi           = {10.1088/0004-637X/806/2/156},
  archiveprefix = {arXiv},
  eprint        = {1504.01203},
  primaryclass  = {astro-ph.GA},
  adsurl        = {https://ui.adsabs.harvard.edu/abs/2015ApJ...806..156S}
}

@article{su17,
  author        = {{Su}, Kung-Yi and {Hopkins}, Philip F. and {Hayward}, Christopher C. and {Faucher-Gigu{\`e}re}, Claude-Andr{\'e} and {Kere{\v{s}}}, Du{\v{s}}an and {Ma}, Xiangcheng and {Robles}, Victor H.},
  title         = {{Feedback first: the surprisingly weak effects of magnetic fields, viscosity, conduction and metal diffusion on sub-L* galaxy formation}},
  journal       = {\mnras},
  year          = 2017,
  month         = oct,
  volume        = {471},
  number        = {1},
  pages         = {144-166},
  doi           = {10.1093/mnras/stx1463},
  archiveprefix = {arXiv},
  eprint        = {1607.05274},
  primaryclass  = {astro-ph.GA},
  adsurl        = {https://ui.adsabs.harvard.edu/abs/2017MNRAS.471..144S}
}

@article{su19,
  author        = {{Su}, Kung-Yi and {Hopkins}, Philip F. and {Hayward}, Christopher C. and {Ma}, Xiangcheng and {Faucher-Gigu{\`e}re}, Claude-Andr{\'e} and {Kere{\v{s}}}, Du{\v{s}}an and {Orr}, Matthew E. and {Chan}, T.~K. and {Robles}, Victor H.},
  title         = {{The failure of stellar feedback, magnetic fields, conduction, and morphological quenching in maintaining red galaxies}},
  journal       = {\mnras},
  year          = 2019,
  month         = aug,
  volume        = {487},
  number        = {3},
  pages         = {4393-4408},
  doi           = {10.1093/mnras/stz1494},
  archiveprefix = {arXiv},
  eprint        = {1809.09120},
  primaryclass  = {astro-ph.GA},
  adsurl        = {https://ui.adsabs.harvard.edu/abs/2019MNRAS.487.4393S}
}

@ARTICLE{su26,
       author = {{Su}, Tingfang and {Ji}, Suoqing and {Yuan}, Feng and {Xia}, Haojie and {Zou}, Yuxuan},
        title = "{Positive AGN Feedback Enhances Star Formation in Starburst Dwarf Galaxies}",
      journal = {\apj},
         year = 2026,
        month = jun,
       volume = {1004},
       number = {1},
          eid = {128},
        pages = {128},
          doi = {10.3847/1538-4357/ae6f12},
archivePrefix = {arXiv},
       eprint = {2510.20897},
 primaryClass = {astro-ph.GA},
       adsurl = {https://ui.adsabs.harvard.edu/abs/2026ApJ..1004..128S}
}

@article{sukhbold16,
  author        = {{Sukhbold}, Tuguldur and {Ertl}, T. and {Woosley}, S.~E. and {Brown}, Justin M. and {Janka}, H. -T.},
  title         = {{Core-collapse Supernovae from 9 to 120 Solar Masses Based on Neutrino-powered Explosions}},
  journal       = {\apj},
  year          = 2016,
  month         = apr,
  volume        = {821},
  number        = {1},
  eid           = {38},
  pages         = {38},
  doi           = {10.3847/0004-637X/821/1/38},
  archiveprefix = {arXiv},
  eprint        = {1510.04643},
  primaryclass  = {astro-ph.HE},
  adsurl        = {https://ui.adsabs.harvard.edu/abs/2016ApJ...821...38S}
}

@article{sun07,
  author        = {{Sun}, M. and {Jones}, C. and {Forman}, W. and {Vikhlinin}, A. and {Donahue}, M. and {Voit}, M.},
  title         = {{X-Ray Thermal Coronae of Galaxies in Hot Clusters: Ubiquity of Embedded Mini-Cooling Cores}},
  journal       = {\apj},
  year          = 2007,
  month         = mar,
  volume        = {657},
  number        = {1},
  pages         = {197-231},
  doi           = {10.1086/510895},
  archiveprefix = {arXiv},
  eprint        = {astro-ph/0606184},
  primaryclass  = {astro-ph},
  adsurl        = {https://ui.adsabs.harvard.edu/abs/2007ApJ...657..197S}
}

@article{sutherland93,
  author   = {{Sutherland}, Ralph S. and {Dopita}, M.~A.},
  title    = {{Cooling Functions for Low-Density Astrophysical Plasmas}},
  journal  = {\apjs},
  year     = 1993,
  month    = sep,
  volume   = {88},
  pages    = {253},
  doi      = {10.1086/191823},
  adsurl   = {https://ui.adsabs.harvard.edu/abs/1993ApJS...88..253S}
}

@article{townsend,
  author        = {{Townsend}, R.~H.~D.},
  title         = {{An Exact Integration Scheme for Radiative Cooling in Hydrodynamical Simulations}},
  journal       = {ApJs},
  year          = 2009,
  month         = apr,
  volume        = {181},
  number        = {2},
  pages         = {391-397},
  doi           = {10.1088/0067-0049/181/2/391},
  archiveprefix = {arXiv},
  eprint        = {0901.3146},
  primaryclass  = {astro-ph.SR},
  adsurl        = {https://ui.adsabs.harvard.edu/abs/2009ApJS..181..391T}
}

@article{vernaleo06,
  author        = {{Vernaleo}, John C. and {Reynolds}, Christopher S.},
  title         = {{AGN Feedback and Cooling Flows: Problems with Simple Hydrodynamic Models}},
  journal       = {\apj},
  year          = 2006,
  month         = jul,
  volume        = {645},
  number        = {1},
  pages         = {83-94},
  doi           = {10.1086/504029},
  archiveprefix = {arXiv},
  eprint        = {astro-ph/0511501},
  primaryclass  = {astro-ph},
  adsurl        = {https://ui.adsabs.harvard.edu/abs/2006ApJ...645...83V}
}

@article{voit05,
  author        = {{Voit}, G. Mark},
  title         = {{Tracing cosmic evolution with clusters of galaxies}},
  journal       = {Reviews of Modern Physics},
  year          = 2005,
  month         = apr,
  volume        = {77},
  number        = {1},
  pages         = {207-258},
  doi           = {10.1103/RevModPhys.77.207},
  archiveprefix = {arXiv},
  eprint        = {astro-ph/0410173},
  primaryclass  = {astro-ph},
  adsurl        = {https://ui.adsabs.harvard.edu/abs/2005RvMP...77..207V}
}

@article{wagner12,
  author        = {{Wagner}, A.~Y. and {Bicknell}, G.~V. and {Umemura}, M.},
  title         = {{Driving Outflows with Relativistic Jets and the Dependence of Active Galactic Nucleus Feedback Efficiency on Interstellar Medium Inhomogeneity}},
  journal       = {\apj},
  year          = 2012,
  month         = oct,
  volume        = {757},
  number        = {2},
  eid           = {136},
  pages         = {136},
  doi           = {10.1088/0004-637X/757/2/136},
  archiveprefix = {arXiv},
  eprint        = {1205.0542},
  primaryclass  = {astro-ph.CO},
  adsurl        = {https://ui.adsabs.harvard.edu/abs/2012ApJ...757..136W}
}

@article{werner19,
  author        = {{Werner}, N. and {McNamara}, B.~R. and {Churazov}, E. and {Scannapieco}, E.},
  title         = {{Hot Atmospheres, Cold Gas, AGN Feedback and the Evolution of Early Type Galaxies: A Topical Perspective}},
  journal       = {\ssr},
  year          = 2019,
  month         = jan,
  volume        = {215},
  number        = {1},
  eid           = {5},
  pages         = {5},
  doi           = {10.1007/s11214-018-0571-9},
  archiveprefix = {arXiv},
  eprint        = {1811.05004},
  primaryclass  = {astro-ph.GA},
  adsurl        = {https://ui.adsabs.harvard.edu/abs/2019SSRv..215....5W}
}

@article{white78,
  author   = {{White}, S.~D.~M. and {Rees}, M.~J.},
  title    = {{Core condensation in heavy halos: a two-stage theory for galaxy formation and clustering.}},
  journal  = {\mnras},
  year     = 1978,
  month    = may,
  volume   = {183},
  pages    = {341-358},
  doi      = {10.1093/mnras/183.3.341},
  adsurl   = {https://ui.adsabs.harvard.edu/abs/1978MNRAS.183..341W}
}

@article{white91,
  author   = {{White}, Simon D.~M. and {Frenk}, Carlos S.},
  title    = {{Galaxy Formation through Hierarchical Clustering}},
  journal  = {\apj},
  year     = 1991,
  month    = sep,
  volume   = {379},
  pages    = {52},
  doi      = {10.1086/170483},
  adsurl   = {https://ui.adsabs.harvard.edu/abs/1991ApJ...379...52W}
}

@article{xie12,
  author        = {{Xie}, Fu-Guo and {Yuan}, Feng},
  title         = {{Radiative efficiency of hot accretion flows}},
  journal       = {\mnras},
  year          = 2012,
  month         = dec,
  volume        = {427},
  number        = {2},
  pages         = {1580-1586},
  doi           = {10.1111/j.1365-2966.2012.22030.x},
  archiveprefix = {arXiv},
  eprint        = {1207.3113},
  primaryclass  = {astro-ph.HE},
  adsurl        = {https://ui.adsabs.harvard.edu/abs/2012MNRAS.427.1580X}
}

@article{yang21,
  author        = {{Yang}, Hai and {Yuan}, Feng and {Yuan}, Ye-Fei and {White}, Christopher J.},
  title         = {{Numerical Simulation of Hot Accretion Flows. IV. Effects of Black Hole Spin and Magnetic Field Strength on the Wind and the Comparison between Wind and Jet Properties}},
  journal       = {\apj},
  year          = 2021,
  month         = jun,
  volume        = {914},
  number        = {2},
  eid           = {131},
  pages         = {131},
  doi           = {10.3847/1538-4357/abfe63},
  archiveprefix = {arXiv},
  eprint        = {2102.03317},
  primaryclass  = {astro-ph.HE},
  adsurl        = {https://ui.adsabs.harvard.edu/abs/2021ApJ...914..131Y}
}

@ARTICLE{2023MNRAS.523..208Y,
       author = {{Yang}, Hai and {Yuan}, Feng and {Kwan}, Tom and {Dai}, Lixin},
        title = "{The properties of wind and jet from a super-Eddington accretion flow around a supermassive black hole}",
      journal = {\mnras},
         year = 2023,
        month = jul,
       volume = {523},
       number = {1},
        pages = {208-220},
          doi = {10.1093/mnras/stad1444},
archivePrefix = {arXiv},
       eprint = {2211.10710},
 primaryClass = {astro-ph.HE},
       adsurl = {https://ui.adsabs.harvard.edu/abs/2023MNRAS.523..208Y}
}

@ARTICLE{2021ARA&A..59..117R,
       author = {{Reynolds}, Christopher S.},
        title = "{Observational Constraints on Black Hole Spin}",
      journal = {\araa},
         year = 2021,
        month = sep,
       volume = {59},
        pages = {117-154},
          doi = {10.1146/annurev-astro-112420-035022},
archivePrefix = {arXiv},
       eprint = {2011.08948},
 primaryClass = {astro-ph.HE},
       adsurl = {https://ui.adsabs.harvard.edu/abs/2021ARA&A..59..117R}
}

@ARTICLE{2019ApJ...873..101Z,
       author = {{Zhang}, Xiaoxia and {Lu}, Youjun},
        title = "{On Constraining the Growth History of Massive Black Holes via Their Distribution on the Spin-Mass Plane}",
      journal = {\apj},
         year = 2019,
        month = mar,
       volume = {873},
       number = {2},
          eid = {101},
        pages = {101},
          doi = {10.3847/1538-4357/ab06c6},
archivePrefix = {arXiv},
       eprint = {1902.07056},
 primaryClass = {astro-ph.HE},
       adsurl = {https://ui.adsabs.harvard.edu/abs/2019ApJ...873..101Z}
}

@article{yoon18,
  author        = {{Yoon}, Doosoo and {Yuan}, Feng and {Gan}, Zhao-Ming and {Ostriker}, Jeremiah P. and {Li}, Ya-Ping and {Ciotti}, Luca},
  title         = {{Active Galactic Nucleus Feedback in an Elliptical Galaxy with the Most Updated AGN Physics. II. High Angular Momentum Case}},
  journal       = {\apj},
  year          = 2018,
  month         = sep,
  volume        = {864},
  number        = {1},
  eid           = {6},
  pages         = {6},
  doi           = {10.3847/1538-4357/aad37e},
  archiveprefix = {arXiv},
  eprint        = {1803.03675},
  primaryclass  = {astro-ph.HE},
  adsurl        = {https://ui.adsabs.harvard.edu/abs/2018ApJ...864....6Y}
}

@article{yuan12,
  author        = {{Yuan}, Feng and {Bu}, Defu and {Wu}, Maochun},
  title         = {{Numerical Simulation of Hot Accretion Flows. II. Nature, Origin, and Properties of Outflows and their Possible Observational Applications}},
  journal       = {\apj},
  year          = 2012,
  month         = dec,
  volume        = {761},
  number        = {2},
  eid           = {130},
  pages         = {130},
  doi           = {10.1088/0004-637X/761/2/130},
  archiveprefix = {arXiv},
  eprint        = {1206.4173},
  primaryclass  = {astro-ph.HE},
  adsurl        = {https://ui.adsabs.harvard.edu/abs/2012ApJ...761..130Y}
}

@article{yuan14,
  author        = {{Yuan}, Feng and {Narayan}, Ramesh},
  title         = {{Hot Accretion Flows Around Black Holes}},
  journal       = {\araa},
  year          = 2014,
  month         = aug,
  volume        = {52},
  pages         = {529-588},
  doi           = {10.1146/annurev-astro-082812-141003},
  archiveprefix = {arXiv},
  eprint        = {1401.0586},
  primaryclass  = {astro-ph.HE},
  adsurl        = {https://ui.adsabs.harvard.edu/abs/2014ARA&A..52..529Y}
}

@article{yuan15,
  author        = {{Yuan}, Feng and {Gan}, Zhaoming and {Narayan}, Ramesh and {Sadowski}, Aleksander and {Bu}, Defu and {Bai}, Xue-Ning},
  title         = {{Numerical Simulation of Hot Accretion Flows. III. Revisiting Wind Properties Using the Trajectory Approach}},
  journal       = {\apj},
  year          = 2015,
  month         = may,
  volume        = {804},
  number        = {2},
  eid           = {101},
  pages         = {101},
  doi           = {10.1088/0004-637X/804/2/101},
  archiveprefix = {arXiv},
  eprint        = {1501.01197},
  primaryclass  = {astro-ph.HE},
  adsurl        = {https://ui.adsabs.harvard.edu/abs/2015ApJ...804..101Y}
}

@article{yuan18,
  author        = {{Yuan}, Feng and {Yoon}, DooSoo and {Li}, Ya-Ping and {Gan}, Zhao-Ming and {Ho}, Luis C. and {Guo}, Fulai},
  title         = {{Active Galactic Nucleus Feedback in an Elliptical Galaxy with the Most Updated AGN Physics. I. Low Angular Momentum Case}},
  journal       = {\apj},
  year          = 2018,
  month         = apr,
  volume        = {857},
  number        = {2},
  eid           = {121},
  pages         = {121},
  doi           = {10.3847/1538-4357/aab8f8},
  archiveprefix = {arXiv},
  eprint        = {1712.04964},
  primaryclass  = {astro-ph.HE},
  adsurl        = {https://ui.adsabs.harvard.edu/abs/2018ApJ...857..121Y}
}

@inproceedings{yuan20,
  author    = {{Yuan}, Feng and {Ostriker}, Jeremiah P. and {Yoon}, DooSoo and {Li}, Ya-Ping and {Ciotti}, Luca and {Gan}, Zhao-Ming and {Ho}, Luis C. and {Guo}, Fulai},
  title     = {{Numerical study of active galactic nucleus feedback in an elliptical galaxy with MACER}},
  booktitle = {Perseus in Sicily: From Black Hole to Cluster Outskirts},
  year      = 2020,
  editor    = {{Asada}, Keiichi and {de Gouveia Dal Pino}, Elisabete and {Giroletti}, Marcello and {Nagai}, Hiroshi and {Nemmen}, Rodrigo},
  series    = {IAU Symposium},
  volume    = {342},
  month     = jan,
  pages     = {101-107},
  doi       = {10.1017/S174392131800371X},
  adsurl    = {https://ui.adsabs.harvard.edu/abs/2020IAUS..342..101Y}
}

@article{zhang25,
  author        = {{Zhang}, Haoen and {Xia}, Haojie and {Ji}, Suoqing and {Yuan}, Feng and {Guo}, Minhang and {Zhang}, Rui and {Zhu}, Bocheng and {Di}, Yihuan and {He}, Aoyun and {Su}, Tingfang and {Zou}, Yuxuan},
  title         = {{MACER3D{\textemdash}An Upgrade of MACER2D with Enhanced Subgrid Models and Gas Physics{\textemdash}and Its Application to Simulating AGN Feedback in a Massive Elliptical Galaxy}},
  journal       = {\apj},
  year          = 2025,
  month         = jun,
  volume        = {985},
  number        = {2},
  eid           = {178},
  pages         = {178},
  doi           = {10.3847/1538-4357/adcaba},
  archiveprefix = {arXiv},
  eprint        = {2504.06342},
  primaryclass  = {astro-ph.GA},
  adsurl        = {https://ui.adsabs.harvard.edu/abs/2025ApJ...985..178Z}
}

@article{zhu23a,
  author        = {{Zhu}, Bocheng and {Yuan}, Feng and {Ji}, Suoqing and {Peng}, Yingjie and {Ho}, Luis C. and {Ostriker}, Jeremiah P. and {Ciotti}, Luca},
  title         = {{Active galactic nuclei feedback in an elliptical galaxy (III): the impacts and fate of cosmological inflow}},
  journal       = {\mnras},
  year          = 2023,
  month         = oct,
  volume        = {524},
  number        = {4},
  pages         = {5787-5803},
  doi           = {10.1093/mnras/stad2055},
  archiveprefix = {arXiv},
  eprint        = {2303.03834},
  primaryclass  = {astro-ph.GA},
  adsurl        = {https://ui.adsabs.harvard.edu/abs/2023MNRAS.524.5787Z}
}

@article{zhu23b,
  author        = {{Zhu}, Bocheng and {Yuan}, Feng and {Ji}, Suoqing and {Peng}, Yingjie and {Ho}, Luis C.},
  title         = {{On the dominant role of wind in the quasar feedback mode in the late-stage evolution of massive elliptical galaxies}},
  journal       = {\mnras},
  year          = 2023,
  month         = nov,
  volume        = {525},
  number        = {4},
  pages         = {4840-4853},
  doi           = {10.1093/mnras/stad2640},
  archiveprefix = {arXiv},
  eprint        = {2308.15970},
  primaryclass  = {astro-ph.GA},
  adsurl        = {https://ui.adsabs.harvard.edu/abs/2023MNRAS.525.4840Z}
}

@ARTICLE{zhu26,
       author = {{Zhu}, Bocheng and {Springel}, Volker and {Yuan}, Feng},
        title = "{Towards physically more comprehensive AGN modelling in cosmological simulations: a MACER-based modification of IllustrisTNG}",
      journal = {\mnras},
         year = 2026,
        month = jun,
       volume = {548},
       number = {4},
          eid = {stag812},
        pages = {stag812},
          doi = {10.1093/mnras/stag812},
archivePrefix = {arXiv},
       eprint = {2603.15235},
 primaryClass = {astro-ph.GA},
       adsurl = {https://ui.adsabs.harvard.edu/abs/2026MNRAS.548ag812Z}
}

@article{zou26a,
  author   = {{Zou}, Yuxuan and {Yuan}, Feng and {Ji}, Suoqing and {Ho}, Luis C. and {Peng}, Yingjie and {Wang}, Jing and {Zhu}, Bocheng and {Wang}, Tao},
  title    = {{A Systematic Study of AGN Feedback in a Disk Galaxy. I. Global Overview}},
  journal  = {\apj},
  year     = 2026,
  month    = mar,
  volume   = {1000},
  number   = {1},
  eid      = {41},
  pages    = {41},
  doi      = {10.3847/1538-4357/ae47fe},
  adsurl   = {https://ui.adsabs.harvard.edu/abs/2026ApJ..1000...41Z}
}

@ARTICLE{zou26b,
       author = {{Zou}, Yuxuan and {Yuan}, Feng and {Ji}, Suoqing and {He}, Lin and {Li}, Zhiyuan and {Zhang}, Yi and {Comparat}, Johan and {Qu}, Zhijie and {Fang}, Taotao},
        title = "{A Systematic Study of AGN Feedback in a Disk Galaxy Using MACER. II. Predictions of X-Ray Surface Brightness Profiles and Comparison with eROSITA Observations}",
      journal = {\apj},
         year = 2026,
        month = jun,
       volume = {1003},
       number = {2},
          eid = {199},
        pages = {199},
          doi = {10.3847/1538-4357/ae6778},
archivePrefix = {arXiv},
       eprint = {2603.22412},
 primaryClass = {astro-ph.GA},
       adsurl = {https://ui.adsabs.harvard.edu/abs/2026ApJ..1003..199Z}
}
    \begin{appendix}
        \section{Comparison of wind-driven and jet-regulated feedback models\label{sec:appendix_a}}

        We briefly summarize the implementation of the jet-included cluster
        simulations in \citet{he26}. Their simulations use the same MACER3D
        framework and the same AGN physics with the present work, except that they
        also include jet. In particular, two models are simulated, namely
        \texttt{JetWind} and \texttt{WindOnly} models. The \texttt{JetWind}
        model includes a collimated jet together with a wide-angle wind, whereas
        their \texttt{WindOnly} model sets the jet mass flux to zero and
        redistributes the jet power into the wind channel. This design enables a
        controlled comparison of feedback kinematics at approximately fixed total
        AGN power.

        In almost all existing sub-grid jet feedback models, the properties of
        jet are usually treated free variables, whose values are not necessarily
        consistent with the constraints of GRMHD simulations of jet formation.
        Different from these works, in \citet{he26}, the jet properties are taken
        from small-scale GRMHD simulations of jet formation \citep[e.g.,][]{yang21}
        or their extrapolation outward to the inner boundary of simulation domain
        ($r_{\rm in}=100\,\mathrm{pc}$) or direct observation. The jet is
        characterized by four key quantities: jet power, bulk velocity, mass flux,
        and half-opening angle. Following their prescription, the total jet
        power is
        \begin{align}
            \dot{E}_{\rm jet}= \dot{E}_{\rm kin,jet}+ \dot{E}_{\rm th,jet}= 0.9\,\dot{M}_{\rm BH}\,c^{2},
        \end{align}
        where $\dot{E}_{\rm kin,jet}$ and $\dot{E}_{\rm th,jet}$ are the kinetic
        and thermal components of the jet power, respectively. The mass flux-weighted
        jet velocity obtained in \citet{yang21} at $200\,r_{g}$ is $0.5\,c$.
        However, the inner boundary of the \citet{he26} simulations are much
        larger than this radius, so the jet velocity at the inner boundary $r_{\rm
        in}$ is taken from observations,
        \begin{align}
            v_{\rm jet}= 0.1\,c.
        \end{align}
        The thermal energy of the jet is correspondingly increased to keep the
        total jet power unchanged. The jet mass flux is obtained from the GRMHD simulation
        presented in \citet{yang21} and is assumed to be remain constant out to
        the inner boundary of the simulation in \citet{he26}, which is
        \begin{align}
            \dot{M}_{\rm jet}= 0.35\,\dot{M}_{\rm BH}.
        \end{align}
        The next jet parameter is the opening angle. \citep{yang21} find that
        the jet radius follows a power-law relation with distance. Using this profile,
        \citet{he26} calculate the half-opening angle of the jet at $10^{4}\,r_{g}$
        is $7.5^{\circ}$. Beyond this radius, up to $10^{7}\,r_{g}$, observations
        indicate that the opening angle of the jet roughly remains constant so
        the jet half-opening angle at the inner boundary is assumed to be
        \begin{align}
            \theta_{\rm j}\approx 7.5^{\circ}.
        \end{align}
        With $\dot{E}_{\rm jet}$, $v_{\rm jet}$, and $\dot{M}_{\rm jet}$
        specified, the kinetic power is determined by
        \begin{align}
            \dot{E}_{\rm kin,jet}= \frac{1}{2}\dot{M}_{\rm jet}v_{\rm jet}^{2},
        \end{align}
        and the remaining power is assigned to the thermal component,
        \begin{align}
            \dot{E}_{\rm th,jet}= \dot{E}_{\rm jet}-\dot{E}_{\rm kin,jet},
        \end{align}
        so that the total injected power is conserved for the adopted fixed
        injection velocity.

        The jet is injected through the inner radial boundary within two polar
        cones ($\theta \le \theta_{\rm j}$ and $\theta \ge \pi-\theta_{\rm j}$).
        In practice, the primitive variables corresponding to the prescribed jet
        mass flux, internal energy, and velocity are imposed in the ghost zones
        at each time step and launched from the injection region at $r_{\rm in}$.
        The jet velocity is taken to be uniform across the jet cross section,
        and no ad hoc jet precession is imposed.

\begin{figure}[!htb]
            \centering
            \includegraphics[width=0.7\columnwidth]{
                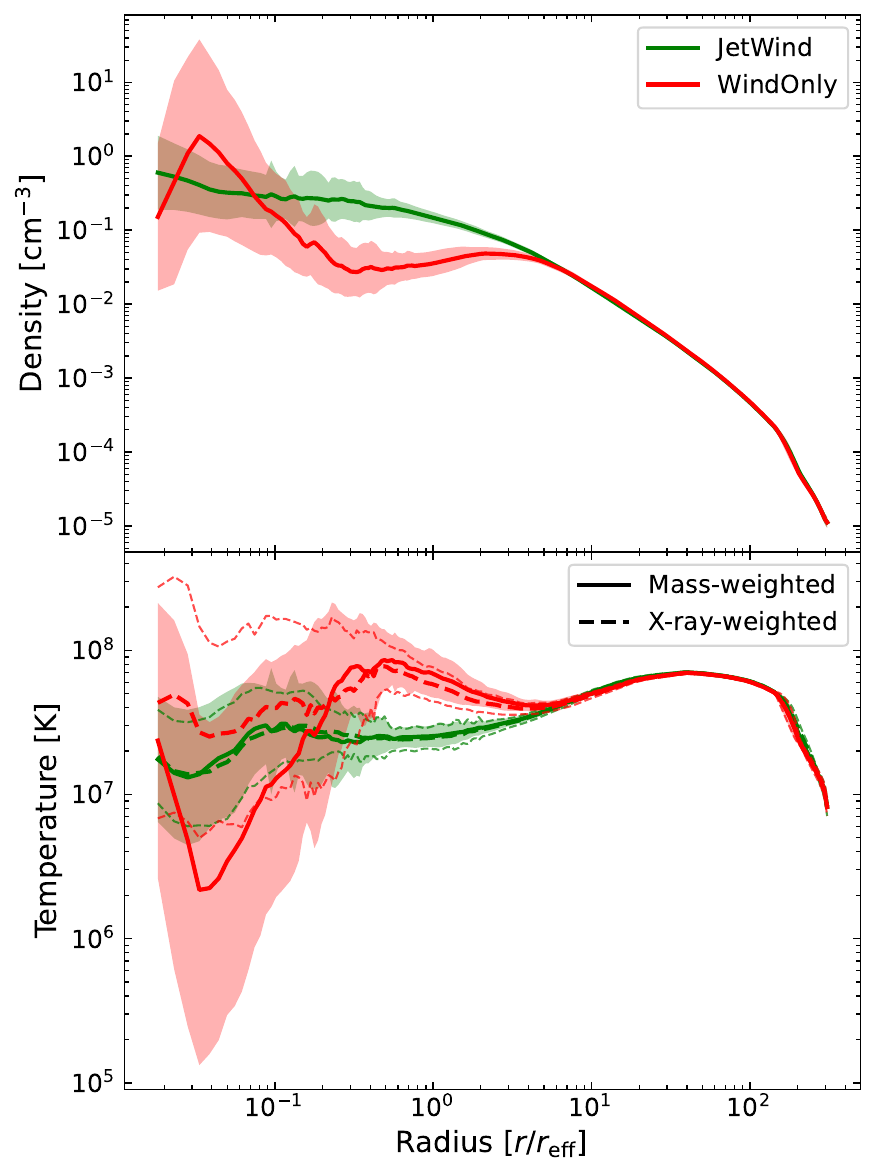
            }
            \caption{Time-averaged radial profiles of the gas properties for the
            cluster models from \citet{he26}. Rows from top to bottom display
            the volume-weighted number density and temperature. Colors denote the
            simulation runs: \texttt{WindOnly} (red) and \texttt{JetWind} (green).
            Shaded regions indicate the temporal scatter. The bottom-row panel
            compares two temperature definitions: mass-weighted (solid) and \textsc{AtomDB}-derived
            X-ray luminosity-weighted (dashed).}
            \label{fig:appendix_profiles}
        \end{figure}

        In addition to the inclusion of jet in \citet{he26}, there are some
        additional setup-level differences between \citet{he26} and our fiducial
        simulations of CCG. \citet{he26} adopt a Perseus-like cool-core cluster setup.
        The total gravitational potential consists of three components: an NFW dark-matter
        halo, the fixed stellar potential of a brightest cluster galaxy (BCG), and a central
        SMBH. They adopt $M_{\rm SMBH}=3.4\times 10^{8}\,M_{\odot}$ and report a
        halo mass of $M_{200}=7.47\times 10^{14}\,M_{\odot}$ (with $r_{200}=1.83\,
        \mathrm{Mpc}$). In their setup, the BCG is NGC~1275, with a stellar mass
        of $M_{\star}=2.43\times 10^{11}\,M_{\odot}$ and an effective radius of $r
        _{\rm eff}=6.41\,\mathrm{kpc}$, comparable to our mE model in stellar scale, while
        the ambient potential is that of a cool-core cluster halo \citep{mathews06}. 
        Although the BCG stellar mass in \citet{he26} is smaller than 
        that of our CCG, the $L_{\rm X}$--$T$ relation is primarily set by the 
        total gravitational potential. The cluster-scale DM halo places their model 
        in a similar temperature regime, making it a useful reference for how jet 
        feedback may affect the hot gas in deep potential wells.
        The ICM is initialized in hydrostatic equilibrium with no imposed rotation
        or perturbations, using observationally constrained radial profiles for
        the electron density and temperature. The simulation domain spans from $r
        _{\rm in}=100\,\mathrm{pc}$ to an outer boundary of $\sim 2\,\mathrm{Mpc}$.
        Their suite includes three models, namely \texttt{WindOnly} (only wind is
        included in the hot mode), \texttt{JetOnly} (only jet is included in the
        hot mode), and \texttt{JetWind} (both jet and wind are included in the hot
        mode). Figure~\ref{fig:appendix_profiles} compares the time-averaged radial
        profiles of the \texttt{WindOnly} and \texttt{JetWind} models from \citet{he26}.

    \onecolumn 
    \section{Supplemental morphological projections for control runs\label{app:control_morphology}}
    We present the full morphological and kinematic snapshots for the two control simulation suites: the \texttt{noAGN} runs and the \texttt{noSN} runs.
    \begin{figure*}[htb]
    \centering
    \includegraphics[width=0.9\textwidth]{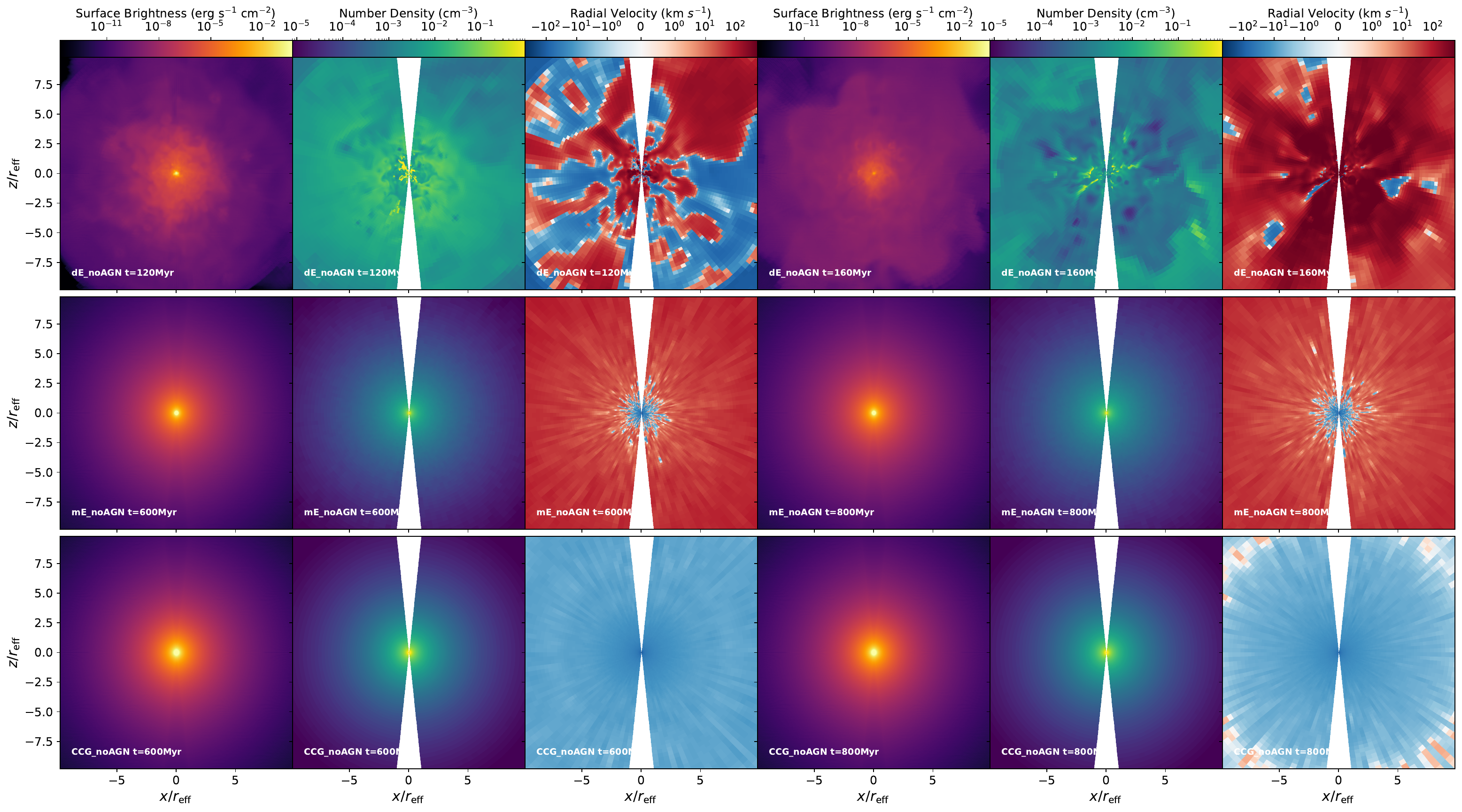}
    \caption{Similar to Fig.~\ref{fig:morphology}, but for \texttt{noAGN} models.}
    \label{fig:morphology_noAGN}
    
    \includegraphics[width=0.9\textwidth]{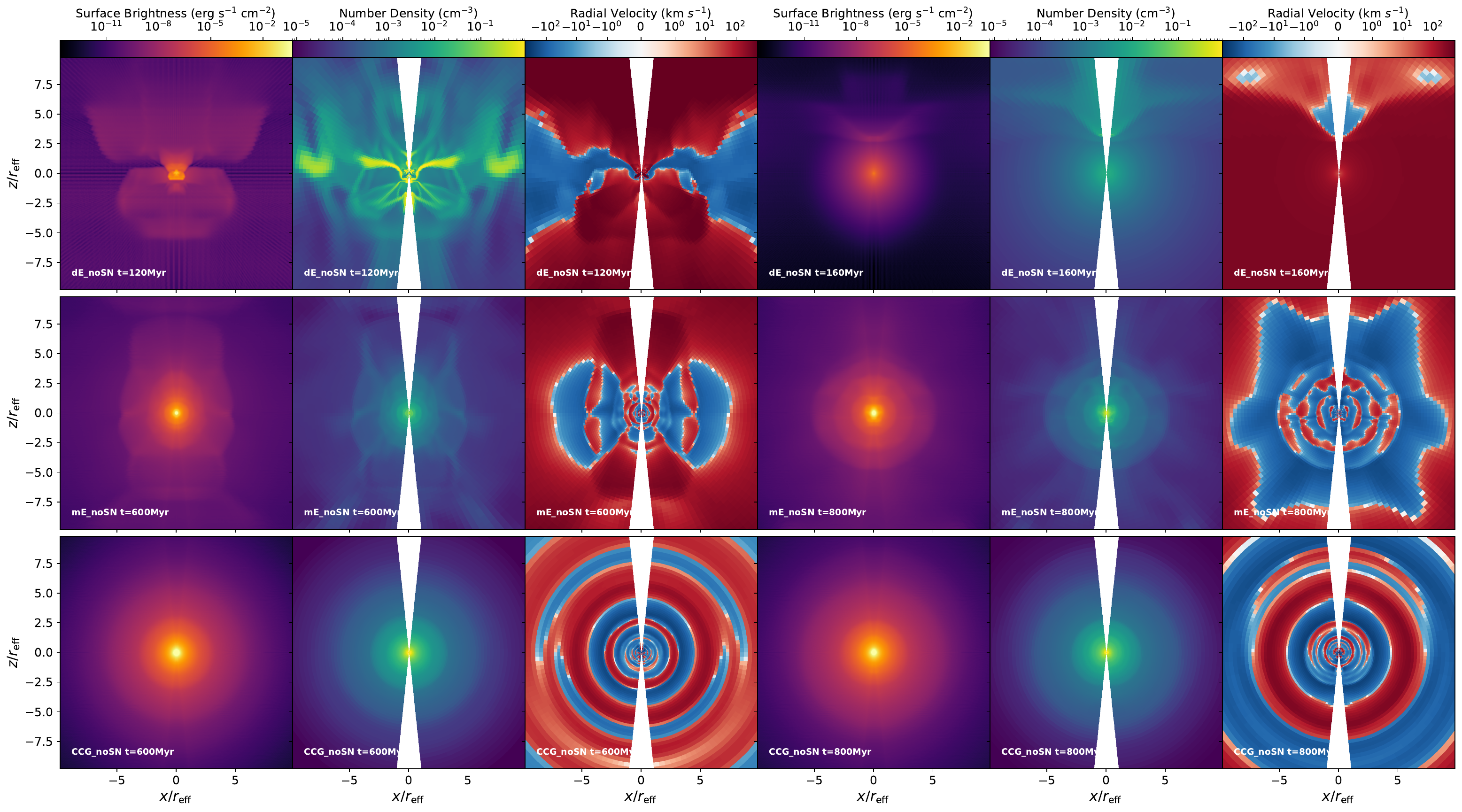}
    \caption{Similar to Fig.~\ref{fig:morphology}, but for \texttt{noSN} models.}
    \label{fig:morphology_noSN}
    \end{figure*}
    
    \end{appendix}
\end{document}